\documentclass[smallextended]{svjour3}       
\smartqed  
\usepackage{graphics,epsfig}
\usepackage{graphicx}
\usepackage{float}
\usepackage{amssymb,amstext,amsmath}
\usepackage{mathtools}
\usepackage{booktabs}
\usepackage{natbib}
\usepackage[utf8]{inputenc}
\usepackage[T1]{fontenc}
\usepackage{epstopdf}
\usepackage{xcolor}
\usepackage{tikz,pgfplots}
\tikzset{
    cross/.pic = {
    \draw[rotate = 45] (-#1,0) -- (#1,0);
    \draw[rotate = 45] (0,-#1) -- (0, #1);
    }
}
\usepackage{physics}
\usepackage{nomencl}
\makenomenclature
\journalname{Archive of Applied Mechanics}

\begin{document}
\newcommand{\balpha}{\boldsymbol{\alpha}}
\newcommand{\bbeta}{\boldsymbol{\beta}}
\newcommand{\bgamma}{\boldsymbol{\gamma}}
\newcommand{\bdelta}{\boldsymbol{\delta}}
\newcommand{\bepsilon}{\boldsymbol{\epsilon}}
\newcommand{\bvarepsilon}{\boldsymbol{\varepsilon}}
\newcommand{\bzeta}{\boldsymbol{\zeta}}
\newcommand{\bfoldeta}{\boldsymbol{\eta}}
\newcommand{\btheta}{\boldsymbol{\theta}}
\newcommand{\bvartheta}{\boldsymbol{\vartheta}}
\newcommand{\biota}{\boldsymbol{\iota}}
\newcommand{\bkappa}{\boldsymbol{\kappa}}
\newcommand{\blambda}{\boldsymbol{\lambda}}
\newcommand{\bmu}{\boldsymbol{\mu}}
\newcommand{\bnu}{\boldsymbol{\nu}}
\newcommand{\bxi}{\boldsymbol{\xi}}
\newcommand{\bpi}{\boldsymbol{\pi}}
\newcommand{\bvarpi}{\boldsymbol{\varpi}}
\newcommand{\brho}{\boldsymbol{\rho}}
\newcommand{\bvarrho}{\boldsymbol{\varrho}}
\newcommand{\bsigma}{\boldsymbol{\sigma}}
\newcommand{\bvarsigma}{\boldsymbol{\varsigma}}
\newcommand{\btau}{\boldsymbol{\tau}}
\newcommand{\bupsilon}{\boldsymbol{\upsilon}}
\newcommand{\bphi}{\boldsymbol{\phi}}
\newcommand{\bvarphi}{\boldsymbol{\varphi}}
\newcommand{\bchi}{\boldsymbol{\chi}}
\newcommand{\bpsi}{\boldsymbol{\psi}}
\newcommand{\bomega}{\boldsymbol{\omega}}
\newcommand{\bGamma}{\boldsymbol{\Gamma}}
\newcommand{\bDelta}{\boldsymbol{\Delta}}
\newcommand{\bTheta}{\boldsymbol{\Theta}}
\newcommand{\bLambda}{\boldsymbol{\Lambda}}
\newcommand{\bXi}{\boldsymbol{\Xi}}
\newcommand{\bPi}{\boldsymbol{\Pi}}
\newcommand{\bSigma}{\boldsymbol{\Sigma}}
\newcommand{\bUpsilon}{\boldsymbol{\Upsilon}}
\newcommand{\bPhi}{\boldsymbol{\Phi}}
\newcommand{\bPsi}{\boldsymbol{\Psi}}
\newcommand{\bOmega}{\boldsymbol{\Omega}}
\def\Xint#1{\mathchoice
   {\XXint\displaystyle\textstyle{#1}}%
   {\XXint\textstyle\scriptstyle{#1}}%
   {\XXint\scriptstyle\scriptscriptstyle{#1}}%
   {\XXint\scriptscriptstyle\scriptscriptstyle{#1}}%
   \!\int}
\def\XXint#1#2#3{{\setbox0=\hbox{$#1{#2#3}{\int}$}
     \vcenter{\hbox{$#2#3$}}\kern-.5\wd0}}
\def\ddashint{\Xint=}
\def\dashint{\Xint-}

\title{Plane constrained shear of single crystals}
\author{F. G\"unther       \and
        K.~C. Le 
}
\institute{F. G\"unther \at
              Lehrstuhl f\"ur Mechanik - Materialtheorie, Ruhr-Universit\"at Bochum, D-44780 Bochum, Germany \\
           \and
           K.~C. Le\,$^{1,2}$ \at
              $^1$\,Materials Mechanics Research Group, Ton Duc Thang University, Ho Chi Minh City, Vietnam
\\
$^2$\,Faculty of Civil Engineering, Ton Duc Thang University, Ho Chi Minh City, Vietnam
\\
 \email{lekhanhchau@tdtu.edu.vn (Corresponding author)}
}

\date{Received: date / Accepted: date}

\maketitle

\begin{abstract}
This paper studies the plane constrained shear problem for single crystals having one active slip system and subjected to loading in both directions within the small strain thermodynamic dislocation theory proposed by Le [2018]. The numerical solution of the boundary value problem shows the combined isotropic and kinematic work hardening, the sensitivity of the stress-strain curves to temperature and strain rate, the Bauschinger effect, and the size effect. 
\end{abstract}

%\mbox{}
\nomenclature{$h,w,L$}{Height, width, and depth of the slab}
\nomenclature{$\vb{s}$}{Unit vector showing the slip direction}
\nomenclature{$\vb{m}$}{Unit vector normal to the slip plane}
\nomenclature{$\varphi$}{Angle between slip direction and $x_1$-axis}
\nomenclature{$\gamma(t)$}{Shear amount (as control parameter)}
\nomenclature{$\gamma_D$}{Energy of one dislocation per unit length}
\nomenclature{$\beta$}{Plastic slip}
\nomenclature{$\vb{u}$}{Displacement vector}
\nomenclature{$\bvarepsilon$}{Total strain tensor}
\nomenclature{$\bvarepsilon^\text{p}$}{Plastic strain tensor}
\nomenclature{$\bvarepsilon^\text{e}$}{Elastic strain tensor}
\nomenclature{$a^2$}{The minimally possible area occupied by one dislocation}
\nomenclature{$b$}{Magnitude of Burgers' vector}
\nomenclature{$\rho$}{Total density of dislocations}
\nomenclature{$\rho^\text{g}$}{Density of non-redundant dislocations}
\nomenclature{$\rho^\text{r}$}{Density of redundant dislocations}
\nomenclature{$\chi$}{Configurational temperature}
\nomenclature{$T$}{Kinetic-vibrational temperature}
\nomenclature{$\bsigma$}{Stress tensor}
\nomenclature{$\tau$}{Resolved shear stress (Schmid stress)}
\nomenclature{$\tau_\text{T}$}{Taylor stress}
\nomenclature{$\tau_\text{Y}$}{Flow stress}
\nomenclature{$\tau_\text{B}$}{Back stress}
\nomenclature{$\mu$}{Shear modulus}
\nomenclature{$\nu$}{Poisson's ratio}
\nomenclature{$T_P$}{Energy barrier expressed in the temperature unit}
\nomenclature{$t_0$}{Time characterizing the depinning rate}
\nomenclature{$q$}{Dimensionless plastic strain rate}
\nomenclature{$q_0$}{Dimensionless total strain rate}
\nomenclature{$\chi_0$}{Steady-state configurational temperature}
\nomenclature{Dot over quantities}{Time rates}
\nomenclature{Bar over quantities}{Quantities averaged over the thickness}
\nomenclature{Tilde over quantities}{Rescaled (dimensionless) quantities}
\printnomenclature

\section{Introduction} \label{S:intro}
When crystalline solids deform, dislocation entanglement together with thermal fluctuation determine the kinetics of dislocation depinning and, thus, the rate of plastic deformation and the isotropic work hardening. In addition, if there are obstacles in form of grain boundaries or precipitates, some of the dislocations, after being depinned and driven by the applied resolved shear stress, may become non-redundant (geometrically necessary) dislocations that pile up near these obstacles giving rise to the non-uniform plastic deformation and the size-dependent kinematic work hardening. Therefore, any plasticity theory aiming at predicting plastic yielding, work hardening, and hysteresis must take the nucleation, multiplication, annihilation, and motion of dislocations into account. The continuum approach to dislocation mediated plasticity is dictated by the high dislocation densities accompanying plastic deformations as well as the disorder induced by the dislocation network. The measure of the latter quantity in terms of the  configurational entropy has been introduced into dislocation mediated plasticity by \citet{langer2010thermodynamic} (see also the earlier work by \citet{berdichevsky2008entropy} where the entropy of microstructure has been defined in a somewhat different way). These authors have formulated two fundamental laws of non-equilibrium thermodynamics applicable to the driven configurational subsystem of dislocations: (i) The first law for the plastic slip rate containing the double exponential function based on the kinetics of dislocation depinning, (ii) The second law necessitating the increase of the configurational entropy toward the maximum achieved at the steady state. It was shown recently by \citet{le2020two,langer2020scaling} that both laws are confirmed by the experiments conducted by \citet{samanta1971dynamic} for copper and aluminum. The so called LBL-theory \citep{langer2010thermodynamic} deduced from these laws predicts correctly the stress-strain curves recorded by \citet{samanta1971dynamic} and \citet{follansbee1988constitutive} during uniform plastic deformations of copper in the wide range of temperatures and strain rates. Its extension that includes thermal softening and adiabatic shear banding, proposed recently in \citep{le2017thermodynamic,le2018thermodynamic}, exhibits quantitative agreement with the experimental observations by \citet{shi1997constitutive}, \citet{abbod2007modeling},  \citet{marchand1988experimental}. The extension of LBL-theory to non-uniform plastic deformation that takes into account the non-redundant (geometrically necessary) dislocations \citep{nye1953some,bilby1955types,kroner1955fundamentale,kroner1958kontinuumstheorie,mura1965continuous,berdichevsky1967dynamic,le1996model,weertman1996dislocation}, called the thermodynamic dislocation theory (TDT), was proposed in \citep{le2018athermodynamic}.  Among various dislocation based plasticity theories we mention here only those in \citep{ortiz1999nonconvex,groma2003spatial,berdichevsky2006continuum,berdichevsky2006thermodynamics,acharya2010new,anand2015stored,levitas2015thermodynamically,hochrainer2016thermodynamically,berdichevsky2019beyond,po2019continuum,lieou2020thermodynamic} which are closely relevant to our thermodynamic approach. 

\citet{le2018athermodynamic} solved the plane constrained shear problem within the small strain TDT approximately by first neglecting the non-redundant dislocations in the loaded specimen. After obtaining the flow stress, the total dislocation density and the configurational temperature, the distribution of non-redundant dislocations appearing in thin boundary layers near the grain boundaries is subsequently found by solving the variational problem similar to that considered in \citep{le2008analytical}. Based on this solution he showed that the stress-strain curves exhibit both the isotropic hardening due to the redundant dislocations and kinematic hardening due to the pile-ups of non-redundant dislocations against the grain boundaries which is size-dependent (see also \citep{berdichevsky2007dislocation,le2008analytical}). In view of the approximate character of this solution, we aim in this paper at clarifying if this behavior is confirmed by the rigorous numerical treatment. Besides, we aim at studying the load reversal leading to the Bauschinger effect as well as its sensitivity with respect to the size of the specimen, the temperature, and the strain rate (cf. \citep{le2018cthermodynamic}).

The paper is organized as follows. The setting of the problem is outlined in Section 2 that contains also the derivation of the governing equations of TDT. Section 3 develops its numerical implementation. In Section 4 we present the results of simulations, the temperature and strain rate sensitivity of the stress-strain curves as well as the size and Bauschinger effects. Finally, Section 5 concludes the paper. 

\section{Plane constrained shear} \label{S:2}

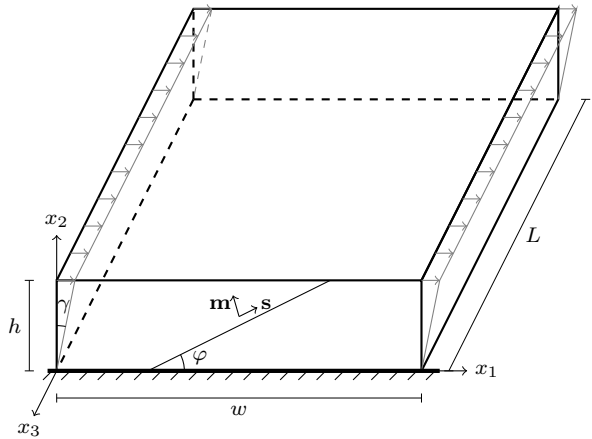
\begin{figure}[hbt!]
\centering
\begin{tikzpicture}[scale=1.2]
\draw[->,thin] (0,0) -- (4.5,0);
\draw[->,thin] (0,0) -- (0,1.5);
\draw[->,thin] (0,0) -- (-0.25,-3/1.5*0.25);
\draw[thick] (0,0) -- (4,0) -- (4,1) -- (0,1) -- (0,0);
\draw[thick] (4,0) -- (5.5,3) -- (5.5,4) -- (4,1);
\draw[thick] (0,1) -- (1.5,4) -- (5.5,4) -- (4,1);
\draw[dashed,thick] (1.5,3) -- (0,0);
\draw[dashed,thick] (1.5,3) -- (1.5,4);
\draw[dashed,thick] (1.5,3) -- (5.5,3);
\draw[gray] (0,0) -- (0.2,1);
\draw[gray] (0.2,1) -- (1.5+0.2,4);
\draw[gray,dashed] (1.5+0.2,4) -- (1.5,3);
\draw[gray] (4,0) -- (4+0.2,1) -- (5.5+0.2,4) -- (5.5,3);
\draw[] (1,0) -- (3,1);
\draw[->] (2,0.6) -- (2.2,0.7);
\draw[->] (2,0.6) -- (1.9333333,0.833333333);
\draw[->,gray] (1.5/10*0,1.5/5*0+1) -- (1.5/10*0+0.2,1.5/5*0+1);
\draw[->,gray] (1.5/10*1,1.5/5*1+1) -- (1.5/10*1+0.2,1.5/5*1+1);
\draw[->,gray] (1.5/10*2,1.5/5*2+1) -- (1.5/10*2+0.2,1.5/5*2+1);
\draw[->,gray] (1.5/10*3,1.5/5*3+1) -- (1.5/10*3+0.2,1.5/5*3+1);
\draw[->,gray] (1.5/10*4,1.5/5*4+1) -- (1.5/10*4+0.2,1.5/5*4+1);
\draw[->,gray] (1.5/10*5,1.5/5*5+1) -- (1.5/10*5+0.2,1.5/5*5+1);
\draw[->,gray] (1.5/10*6,1.5/5*6+1) -- (1.5/10*6+0.2,1.5/5*6+1);
\draw[->,gray] (1.5/10*7,1.5/5*7+1) -- (1.5/10*7+0.2,1.5/5*7+1);
\draw[->,gray] (1.5/10*8,1.5/5*8+1) -- (1.5/10*8+0.2,1.5/5*8+1);
\draw[->,gray] (1.5/10*9,1.5/5*9+1) -- (1.5/10*9+0.2,1.5/5*9+1);
\draw[->,gray] (1.5/10*10,1.5/5*10+1) -- (1.5/10*10+0.2,1.5/5*10+1);
\draw[->,gray] (1.5/10*0+4,1.5/5*0+1) -- (1.5/10*0+0.2+4,1.5/5*0+1);
\draw[->,gray] (1.5/10*1+4,1.5/5*1+1) -- (1.5/10*1+0.2+4,1.5/5*1+1);
\draw[->,gray] (1.5/10*2+4,1.5/5*2+1) -- (1.5/10*2+0.2+4,1.5/5*2+1);
\draw[->,gray] (1.5/10*3+4,1.5/5*3+1) -- (1.5/10*3+0.2+4,1.5/5*3+1);
\draw[->,gray] (1.5/10*4+4,1.5/5*4+1) -- (1.5/10*4+0.2+4,1.5/5*4+1);
\draw[->,gray] (1.5/10*5+4,1.5/5*5+1) -- (1.5/10*5+0.2+4,1.5/5*5+1);
\draw[->,gray] (1.5/10*6+4,1.5/5*6+1) -- (1.5/10*6+0.2+4,1.5/5*6+1);
\draw[->,gray] (1.5/10*7+4,1.5/5*7+1) -- (1.5/10*7+0.2+4,1.5/5*7+1);
\draw[->,gray] (1.5/10*8+4,1.5/5*8+1) -- (1.5/10*8+0.2+4,1.5/5*8+1);
\draw[->,gray] (1.5/10*9+4,1.5/5*9+1) -- (1.5/10*9+0.2+4,1.5/5*9+1);
\draw[->,gray] (1.5/10*10+4,1.5/5*10+1) -- (1.5/10*10+0.2+4,1.5/5*10+1);
\draw[ultra thin] (-0.3,0) -- (-0.3,1);
\draw[ultra thin] (-0.3+0.05,0) -- (-0.3-0.05,0);
\draw[ultra thin] (-0.3+0.05,1) -- (-0.3-0.05,1);
\draw[ultra thin] (0,-0.3) -- (4,-0.3);
\draw[ultra thin] (0,-0.3+0.05) -- (0,-0.3-0.05);
\draw[ultra thin] (4,-0.3+0.05) -- (4,-0.3-0.05);
\draw[ultra thin] (4.3,0) -- (5.8,3);
\draw[ultra thin] (4.3+0.05,0) -- (4.3-0.05,0);
\draw[ultra thin] (5.8+0.05,3) -- (5.8-0.05,3);
\draw[ultra thick] (-0.1,0) -- (4.2,0);
\draw[thin] (-0.05+4/20*0,0) -- (-0.15+4/20*0,-0.1);
\draw[thin] (-0.05+4/20*1,0) -- (-0.15+4/20*1,-0.1);
\draw[thin] (-0.05+4/20*2,0) -- (-0.15+4/20*2,-0.1);
\draw[thin] (-0.05+4/20*3,0) -- (-0.15+4/20*3,-0.1);
\draw[thin] (-0.05+4/20*4,0) -- (-0.15+4/20*4,-0.1);
\draw[thin] (-0.05+4/20*5,0) -- (-0.15+4/20*5,-0.1);
\draw[thin] (-0.05+4/20*6,0) -- (-0.15+4/20*6,-0.1);
\draw[thin] (-0.05+4/20*7,0) -- (-0.15+4/20*7,-0.1);
\draw[thin] (-0.05+4/20*8,0) -- (-0.15+4/20*8,-0.1);
\draw[thin] (-0.05+4/20*9,0) -- (-0.15+4/20*9,-0.1);
\draw[thin] (-0.05+4/20*10,0) -- (-0.15+4/20*10,-0.1);
\draw[thin] (-0.05+4/20*11,0) -- (-0.15+4/20*11,-0.1);
\draw[thin] (-0.05+4/20*12,0) -- (-0.15+4/20*12,-0.1);
\draw[thin] (-0.05+4/20*13,0) -- (-0.15+4/20*13,-0.1);
\draw[thin] (-0.05+4/20*14,0) -- (-0.15+4/20*14,-0.1);
\draw[thin] (-0.05+4/20*15,0) -- (-0.15+4/20*15,-0.1);
\draw[thin] (-0.05+4/20*16,0) -- (-0.15+4/20*16,-0.1);
\draw[thin] (-0.05+4/20*17,0) -- (-0.15+4/20*17,-0.1);
\draw[thin] (-0.05+4/20*18,0) -- (-0.15+4/20*18,-0.1);
\draw[thin] (-0.05+4/20*19,0) -- (-0.15+4/20*19,-0.1);
\draw[thin] (-0.05+4/20*20,0) -- (-0.15+4/20*20,-0.1);
\draw[thin] (-0.05+4/20*21,0) -- (-0.15+4/20*21,-0.1);
\draw (1.4,0) arc[radius = 0.4, start angle= 0, end angle= 26.565];
\draw (0,0.5) arc[radius = 0.5, start angle= 90, end angle= 78.69];
\path[] (1.4,0.15) node[right] {\footnotesize $\varphi$};
\path[] (0.08,0.55) node[above] {\footnotesize $\gamma$};
\path[] (4.5,0) node[right] {\footnotesize $x_1$};
\path[] (0,1.5) node[above] {\footnotesize $x_2$};
\path[] (-0.3,-0.5) node[below] {\footnotesize $x_3$};
\path[] (2.3,0.6) node[above] {\footnotesize $\vb{s}$};
\path[] (2,0.733333333) node[left] {\footnotesize $\vb{m}$};
\path[] (-0.3,0.5) node[left] {\footnotesize $h$};
\path[] (2,-0.3) node[below] {\footnotesize $w$};
\path[] (5.05,1.5) node[right] {\footnotesize $L$};
\end{tikzpicture}
\caption{Single crystal subjected to the time-dependent plane constrained shear $\gamma(t)$ with an active slip system inclined under the angle $\varphi$.}
\label{simpleshear}
\end{figure}

Let a thin slab, made of a single crystal, with width $w$, height $h$ and depth $L$, where $0\leq x_1\leq w$, $0\leq x_2\leq h$ and $-L\leq x_3\leq 0$, be subjected to a shear controlled test (see Fig.~\ref{simpleshear}). We assume that the depth of the slab is the dominant length scale, while the height of the slab is much smaller than its width $(L\gg w\gg h)$, so that the boundary effects can be neglected for $x_1=0$ and $x_1=w$. Based on this assumption, the independent variables are reduced to the spatial coordinate $x_2$ and the time $t$. Besides, only one active slip system is admitted, whose slip direction $\vb{s}$ forms the angle $\varphi$ with the $x_1$-axis, while the dislocation lines are oriented parallel to the $x_3$-axis.

The slab is clamped on the lower side and deformed on the upper side with the given shear $\gamma(t)$, so that the temporal development of $\gamma(t)$ evokes the changing load. Thus the kinematic boundary conditions are 
\begin{equation}
u_1(0,t)=0,\quad u_2(0,t)=0,\quad u_1(h,t)=\gamma(t)h,\quad u_2(h,t)=0,
\label{kinematic_bc}
\end{equation}
with $u_1(x_2,t)$ and $u_2(x_2,t)$ being the non-zero components of the displacement vector $\vb{u}(x_2,t)$ ($u_3\equiv 0$). As these hard conditions do not allow dislocations to reach the upper and lower boundaries, we set
\begin{equation}
\beta(0,t)=0,\quad\beta(h,t)=0
\label{plasticslip_bc}
\end{equation} 
for the plastic slip $\beta(x_2,t)$. Thus, the lower and upper sides act as grain boundaries that hinder the upward and downward movement of edge dislocations. 

For the underlying plane shear, the total strain tensor $\bvarepsilon=\frac{1}{2}(\grad \vb{u}+\vb{u}\grad )$ takes the form
\begin{equation}
\bvarepsilon=\frac{1}{2}\left(\begin{matrix}
0 & u_{1,2} & 0\\
u_{1,2} & 2u_{2,2} & 0\\ 
0 & 0 & 0
\end{matrix}\right) ,
%\label{strain}
\end{equation}
where the comma denotes the derivative with respect to $x_2$. The active slip system is characterized by two unit vectors, where $\vb{s}$ indicates the slip direction and $\vb{m}$ the normal to the slip plane. They are given by
\begin{equation}
\vb{s}=\left(\begin{matrix}
\cos \varphi \\
\sin \varphi \\
0
\end{matrix}\right),\quad \vb{m}=\left(\begin{matrix}
-\sin \varphi \\
\cos \varphi \\
0
\end{matrix}\right) .
%\label{slipsystem}
\end{equation}
In terms of these vectors the plastic distortion can be written as $\bbeta=\beta(x_2,t) \vb{s}\otimes \vb{m}$. In matrix form this tensor equation reads
\begin{equation}
\bbeta=\beta(x_2,t)\left(\begin{matrix}
-\sin\varphi \cos\varphi & \cos^2\varphi & 0\\
-\sin^2\varphi & \sin\varphi\cos\varphi & 0\\
0 & 0 & 0
\end{matrix}\right) .
%\label{plasticdistortion}
\end{equation}
The plastic strain tensor $\bvarepsilon^\text{p}=\frac{1}{2}(\bbeta+\bbeta^T)$ equals 
\begin{equation}
\bvarepsilon^\text{p}=\frac{1}{2}\beta(x_2,t)\left(\begin{matrix}
-\sin2\varphi & \cos2\varphi & 0\\
\cos2\varphi & \sin2\varphi & 0\\
0 & 0 & 0
\end{matrix}\right).
%\label{plastic_strain}
\end{equation}
Furthermore, the elastic strain tensor $\bvarepsilon^\text{e}=\bvarepsilon-\bvarepsilon^{p}$ is given by  
\begin{equation}
\bvarepsilon^\text{e}=\frac{1}{2}\left(\begin{matrix}
\beta\sin2\varphi & u_{1,2}-\beta\cos2\varphi & 0\\
u_{1,2}-\beta\cos2\varphi & 2u_{2,2}-\beta\sin2\varphi & 0\\
0 & 0 & 0
\end{matrix}\right),
%\label{elastic_strain}
\end{equation}
while for the Nye's dislocation density tensor $\balpha=-\bbeta\cross\grad$ we have
\begin{equation}
\balpha=\beta_{,2} \sin\varphi \left(\begin{matrix}
0 & 0 & \cos\varphi\\
0 & 0 & \sin\varphi\\ 
0 & 0 & 0
\end{matrix}\right).
%\label{Nyetensor}
\end{equation}
With this the scalar density of non-redundant dislocations per unit area perpendicular to the $x_3$-axis is quantified according to
\begin{equation}
\rho^\text{g}=\frac{1}{b}\vert \balpha \vdot \vb{e}_3 \vert =\frac{1}{b}\vert\beta_{,2}\sin\varphi \vert,
%\label{scalar_density}
\end{equation}
where $b$ is the magnitude of the Burgers' vector. Note that $\rho^\text{g}$ can be measured by the high-resolution EBSD-technique (see, e.g., \citep{calcagnotto2010orientation}). Since the total dislocation density $\rho$ can be measured with the TEM \citep{morito2003dislocation} or the XRD-technique \citep{ayers1994measurment}, the density of the redundant dislocation $\rho^\text{r}=\rho-\rho^\text{g}$ can in principle also be measured.

With regard to this two-dimensional problem the energy functional per unit depth from \citep{le2018athermodynamic} takes the following form
\begin{align}
I\left[u_1,u_2,\beta,\rho^\text{r},\chi\right]= w \int_0^h\Bigl[\frac{1}{2}\lambda u_{2,2}^2+\frac{1}{2}\mu\left(u_{1,2}-\beta\cos2\varphi\right)^2\notag\\
+\frac{1}{4}\mu\beta^2\sin^2 2\varphi+\mu\left(u_{2,2}-\frac{1}{2}\beta\sin2\varphi\right)^2+\gamma_\text{D}\rho^\text{r}\notag\\
+\gamma_\text{D}\rho^\text{s}\ln\Bigl(\frac{1}{1-\frac{1}{\rho^\text{s} b}\vert\beta_{,2}\sin\varphi\vert}\Bigr)-\frac{\chi}{L} \left(-\rho\ln\left(a^2\rho\right)+\rho\right) \Bigr] \dd{x_2}.
\label{energiefunctional}
\end{align}
The four first terms in \eqref{energiefunctional} describe energy of crystal due to the elastic strain, with $\mu$ and $\lambda$ being Lam\'{e} constants (for simplicity of the analysis, the crystal is assumed to be elastically isotropic). The fifth term is the self-energy of redundant (statistically stored) dislocations, with $\gamma_\text{D}$ being the energy of one dislocation per unit length. The sixth term is the energy of non-redundant dislocations, where $\rho^s$ denotes a saturated dislocation density \citep{berdichevsky2006thermodynamics}. The last term has been introduced by \citet{langer2015}, with $S_\text{C} = (-\rho \ln(a^2 \rho) + \rho)$ being the configurational entropy of dislocations per unit area, $a^2$ the minimally possible area occupied by one dislocation, and $\chi/L$ the ``two-dimensional''  configurational temperature. For $S_\text{C}=\partial F/\partial \chi$ to be a function of $\rho$ and $\chi$, we assume that $a$ is a slowly increasing function of $\chi$ \citep{le2020introduction}. Note that \citet{berdichevsky2005homogenization} has calculated a similar term for anti-plane shear. However, his result cannot be applied here for two reasons: (i) In \citep{berdichevsky2005homogenization} only screw dislocations of the same sign are considered, (ii) the loading is assumed to be quasi-static. By varying the functional with respect to $u_1$ and $u_2$ and integrating the resulting equations with the use of the boundary conditions \eqref{kinematic_bc}, a reduction of the arguments of energy to only $\beta$, $\rho^\text{r}$ and $\chi$ can be achieved with 
\begin{equation}
u_{1,2}=\gamma+(\beta-\langle\beta\rangle)\cos2\varphi,\qquad u_{2,2}=\kappa(\beta-\langle\beta\rangle)\sin2\varphi ,
%\label{integrated_equilibrium}
\end{equation}
where $\kappa =\frac{\mu }{\lambda +2\mu }$, and $\langle \beta \rangle =\frac{1}{h}\int_0^h\beta \dd{x_2}$. Inserting the two equations for $u_{1,2}$ and $u_{2,2}$ into the energy functional \eqref{energiefunctional} yields its relaxed form
\begin{multline}
I= w \int_0^h\Bigl[\frac{1}{2}\mu\kappa\langle\beta\rangle^2\sin^2 2\varphi+\frac{1}{2}\mu(\langle\beta\rangle\cos2\varphi-\gamma)^2+\frac{1}{2}\mu(1-\kappa)\beta^2\sin^2 2\varphi\\
+\gamma_\text{D}\rho^\text{r}
+\gamma_\text{D}\rho^\text{s}\ln\Bigl( \frac{1}{1-\frac{1}{\rho^\text{s}b}\vert\beta_{,2}\sin\varphi \vert}\Bigr)
-\frac{\chi}{L}\Bigl(-\rho\ln (a^2\rho)+\rho\Bigr) \Bigr] \dd{x_2}.
\label{relaxed_energie}
\end{multline}

In addition to the energy, the dissipation potential must also be proposed. According to \citep{le2018athermodynamic} we take it in the form  
\begin{equation}
D(\dot{\beta},\dot{\rho},\dot{\chi})=\tau _Y \dot{\beta }+\frac{1}{2}d_\rho \dot{\rho }^2+\frac{1}{2}d_\chi \dot{\chi }^2,
\end{equation}
where $\tau_Y$ is the flow stress, $d_\rho $ and $d_\chi $ need be determined so that the governing equations are reduced to those of the LBL-theory for uniform plastic deformation. We formulate the following variational principle \citep{le2018athermodynamic}: the true plastic slips $\check{\beta }(x_2,t)$, the true density of redundant dislocations $\check{\rho }^\text{r}(x_2,t)$, and the true configurational temperature $\check{\chi }(x_2,t)$ obey the variational equation
\begin{equation}
\delta I+w\int_{0}^h \Bigl( \frac{\partial D}{\partial \dot{\beta }}\delta \beta +\frac{\partial D}{\partial \dot{\rho }}\delta \rho +\frac{\partial D}{\partial \dot{\chi }}\delta \chi \, \Bigr) \dd{x_2}=0
\end{equation}
for all variations of admissible fields $\beta (x_2,t)$, $\rho ^\text{r}(x_2,t)$, and $\chi (x_2,t)$ satisfying the constraints \eqref{plasticslip_bc}.

For the considered problem of plane constrained shear the evolution equations of TDT for $\beta$, $\rho^\text{r}$ and $\chi$ read
\begin{align}
\langle\dot{\beta}\rangle&=\dfrac{q(\tau_\text{Y},\rho^\text{r},T)}{t_0},\quad q\left(\tau_\text{Y},\rho^\text{r},T\right)=b\sqrt{\rho^\text{r}}\left[ f_\text{P}\left(\tau_\text{Y},\rho^\text{r},T\right)-f_\text{P}\left(-\tau_\text{Y},\rho^\text{r},T\right)\right], \notag
\\
\dot{\rho}&=\dfrac{\mathcal{K}_{\rho}}{a^2\mu\zeta(\rho^\text{r},q_0,T)^2}\tau\dfrac{q\left(\tau_\text{Y},\rho^\text{r},T\right)}{t_0}\left(1-\frac{\rho}{\rho^\text{s}(\chi)}\right), \label{eq_beta_rho_chi}
\\
\dot{\chi}&=\frac{\mathcal{K}_\chi}{\mu}\tau\frac{q\left(\tau_\text{Y},\rho^\text{r},T\right)}{t_0}\left(1-\frac{\chi}{\chi_0}\right). \notag
\end{align}
Here $t_0$ is the time characterizing the depinning rate, $T$ the ordinary temperature, $\tau_\text{T}(\rho^\text{r})=\mu_\text{T}b\sqrt{\rho^\text{r}}$ - the Taylor stress,
\begin{equation}
%\label{f_P}
f_\text{P}\left(\tau_\text{Y},\rho^\text{r},T\right)\equiv \exp \Bigl[-\frac{T_\text{P}}{T} e^{-\tau_\text{Y}/\tau_\text{T}(\rho^\text{r})} \Bigr] 
\end{equation}
is the double exponential function originating from the kinetics of dislocation depinning, while the double logarithmic function
\begin{equation}
%\label{nu}
\zeta(\rho^\text{r},q_0,T)=\ln \Bigl( \frac{T_\text{P}}{T}\Bigr)-\ln\Bigl[\ln\Bigl(\frac{b\sqrt{\rho^\text{r}}}{q_0}\Bigr)\Bigr] 
\end{equation}
has the meaning of the stress ratio $\tau_\text{Y}/\tau_\text{T}$ \citep{langer2010thermodynamic}. Note that, when dealing with the load reversal, antisymmetry is required in Eq.~\eqref{eq_beta_rho_chi}$_1$ for $q(\tau_\text{Y},\rho^\text{r},T)$ both to preserve reflection symmetry, and to satisfy the second-law requirement that the energy dissipation rate, $\tau_Yq/q_0$, is non-negative. In contrary, in the balance of microforces acting on non-redundant dislocations
\begin{equation}
\tau-\tau_\text{B}-\tau_\text{Y}=0
%\label{microforce}
\end{equation}
obtained by varying \eqref{relaxed_energie} with respect to $\beta$, both the resolved shear stress (Schmid stress) $\tau=\vb{s}\vdot \bsigma \vdot \vb{m}$ and the back stress $\tau_\text{B}$ must be 
\begin{align}
\tau&=-\mu\left(\kappa\langle\beta\rangle\sin^2 2\varphi+(\langle\beta\rangle\cos 2\varphi-\gamma)\cos2\varphi+(1-\kappa)\beta\sin^2 2\varphi \right), \label{schmidstress}
\\
\tau_\text{B}&=-\dfrac{C_1}{\bigl(1-C_2\vert\beta_{,2}\vert\bigr)^2}\beta_{,22},\quad\text{$C_1=\frac{\gamma_\text{D}}{\rho^\text{s}b^2}\sin^2\varphi ,\quad C_2=\frac{1}{\rho^\text{s}b}\vert\sin\varphi\vert$}.
\label{backstress}
\end{align}
Note that the back stress $\tau_B$, obtained by varying the energy term containing $\beta_{,2}$ and integrating by parts using the kinematic boundary condition \eqref{plasticslip_bc}, describes the interaction between non-redundant dislocations of equal sign. To derive the evolution equation for the flow stress $\tau_\text{Y}$ let us consider first the uniform total and plastic deformations for which $\tau_\text{Y}=\tau$. Taking the time derivative of $\tau$ from \eqref{schmidstress} we get
\begin{equation}
\dot{\tau}=-\mu\left(\kappa\langle\dot{\beta}\rangle\sin^2 2\varphi+(\langle\dot{\beta}\rangle\cos2\varphi-\dot{\gamma})\cos2\varphi+(1-\kappa)\dot{\beta}\sin^2 2\varphi \right).
%\label{stress_rate}
\end{equation}
Using the evolution equation \eqref{eq_beta_rho_chi}$_1$, we obtain for $\dot{\tau}_\text{Y}$ 
\begin{equation}
\dot{\tau}_\text{Y}=\mu\Bigl(\dot{\gamma}\cos2\varphi-\frac{q}{t_0}\Bigr) .
\label{evolution_yield}
\end{equation}
As the flow stress $\tau_\text{Y}$ determines the overall dislocation depinning process, we postulate that \eqref{evolution_yield} is fulfilled in the most general case of nonuniform plastic deformations. To obtain the system of equations directly in term of the changing shear strain, a constant shear rate $\dot{\gamma}=q_0/t_0$ is assumed as \citet{langer2010thermodynamic} did. In this case, the time rate in the system of PDEs can be replaced by the rate with respect to the total amount of shear $\gamma$ according to the relation $t_0\partial /\partial t=q_0\partial /\partial \gamma$, whereby the evolution equation for the average plastic slip can be transformed to 
\begin{equation}
\pdv{\langle \beta \rangle}{\gamma}=\frac{q(\tau_\text{Y},\rho^\text{r},T)}{q_0}.
%\label{dbeta_dgamma}
\end{equation}
The final system of PDEs governing the evolution of loaded crystal reads:
\begin{align}
&\pdv{\tau_\text{Y}}{\gamma}=\mu\Bigl(\cos2\varphi -\frac{q(\tau_\text{Y},\rho^\text{r},T)}{q_0}\Bigr),\notag\\
&\pdv{\chi}{\gamma}=\frac{\mathcal{K}_\chi}{\mu_\text{T}}\tau\frac{q(\tau_\text{Y},\rho^\text{r},T)}{q_0}\Bigl(1-\dfrac{\chi}{\chi_0}\Bigr),\label{final_equations}
\\
&\pdv{\rho}{\gamma}=\frac{\mathcal{K}_{\rho}}{a^2\mu\zeta(\rho^\text{r},q_0,T)^2}\tau\frac{q(\tau_\text{Y},\rho^\text{r},T)}{q_0}\Bigl(1-\frac{\rho}{\rho^\text{s}(\chi)}\Bigr),\notag\\
&\tau-\tau_\text{B}-\tau_\text{Y}=0. \notag
\end{align}
These equations are subjected to the initial and boundary conditions \eqref{plasticslip_bc}.

\section{Numerical implementation} \label{S:3}
In the previous Section the PDEs governing the plane constrained shear deformation of single crystal have been derived. As mentioned in the Introduction, the approximate solution of this system has been found in \citep{le2018athermodynamic}. With the aim of verifying the obtained result, the numerical solution algorithm of these PDEs based on the finite difference method is developed in the present Section (cf. also \citep{le2018cthermodynamic,le2019thermodynamic}).

First, for the numerical integration of system \eqref{final_equations}, it is convenient to use the rescaled variables and unknown functions according to 
\begin{equation}
\tilde{x}_2=\frac{x_2}{b},\quad\tilde{\rho}=a^2\rho,\quad\tilde{\chi}=\frac{\chi}{e_\text{D}},\quad\tilde{\tau}=\frac{\tau}{\mu},\quad\tilde{\tau}_\text{Y}=\frac{\tau_\text{Y}}{\mu},\quad\tilde{\tau}_\text{B}=\frac{\tau_\text{B}}{\mu}.
%\label{rescaled}
\end{equation}
In terms of these variables and unknown functions $\tilde{\rho}^\text{g}=|\beta_{,\tilde{2}}\sin \varphi|$, while $\tilde{\rho}^\text{r}=\tilde{\rho}-|\beta_{,\tilde{2}}\sin \varphi|$. If $\tilde{\mu}_\text{T}=(b/a)\mu_\text{T}=\mu r$ is used, with $r$ being a dimensionless quantity independent of the loading rate as well as the ordinary temperature, and the dimensionless ordinary temperature $\theta$ is defined as the ratio between $T$ and the activation temperature 
\begin{equation}
\theta=\frac{T}{T_\text{P}},
%\label{dimensionles_temperature}
\end{equation}
then the dimensionless plastic slip rate can be rewritten as
\begin{equation}
q\left(\tau_\text{Y},\rho^\text{r},T\right)=\frac{b}{a}\tilde{q}(\tilde{\tau}_\text{Y},\tilde{\rho}^\text{r},\theta),\quad \tilde{q}(\tilde{\tau}_\text{Y},\tilde{\rho}^\text{r},\theta)=\sqrt{\tilde{\rho}^\text{r}}[\tilde{f}_\text{P}(\tilde{\tau}_\text{Y},\tilde{\rho}^\text{r},\theta)-\tilde{f}_\text{P}(-\tilde{\tau}_\text{Y},\tilde{\rho}^\text{r},\theta)],
\label{dimensionless_plastic_sliprate}
\end{equation}
where
\begin{equation}
\tilde{f}_\text{P}\left(\tilde{\tau}_\text{Y},\tilde{\rho},\theta\right)=\exp\Bigl[-\frac{1}{\theta}\exp\Bigl(-\frac{\tilde{\tau}_\text{Y}}{r\sqrt{\tilde{\rho}^\text{r}}}\Bigr)\Bigr].
%\label{dimensionless_f_P}
\end{equation}
For the target steady-state dislocation density $\rho^\text{s}(\chi)=(1/a^2)e^{-e_\text{D}/\chi}$ and configurational temperature $\chi$, the dimensionless quantities
\begin{equation}
\tilde{\rho}^\text{s}(\tilde{\chi})=\exp\Bigl(-\frac{1}{\tilde{\chi}}\Bigr),\quad\tilde{\chi}_0=\frac{\chi_0}{e_\text{D}}
%\label{steady_state}
\end{equation}
should be used. Note that the saturated dislocation density is assumed to be equal to the steady-state dislocation density at the maximum configurational temperature $\chi=\chi_0$: $\rho^\text{s}=\rho^\text{s}(\chi_0)=(1/a^2)e^{-e_\text{D}/\chi_0}$. The dimensionless plastic slip rate $\tilde{q}$ effectively leads to a rescaling of the time $t_0$ by the factor $b/a$. Following the suggestion made by \citet{langer2010thermodynamic}, $\tilde{t}_0=(a/b) t_0=10^{-12}\text{s}$ is assumed. Correspondingly, the shear rate $\dot{\gamma}=\tilde{q}_0/\tilde{t}_0$.

The set of governing PDEs of the present material model in the dimensionless form more accessible for numerical integration is thus 
\begin{align}
&\pdv{\tilde{\tau}_\text{Y}}{\gamma}=\cos2\varphi-\frac{\tilde{q}(\tilde{\tau}_\text{Y},\tilde{\rho}^\text{r},\theta)}{\tilde{q}_0},\notag\\
&\pdv{\tilde{\chi}}{\gamma}=\mathcal{K}_\chi\tilde{\tau}_\text{Y}\frac{\tilde{q}(\tilde{\tau}_\text{Y},\tilde{\rho}^\text{r},\theta)}{\tilde{q}_0}\Bigl(1-\frac{\tilde{\chi}}{\tilde{\chi}_0}\Bigr),\label{dimensionless_PDE}\\
&\pdv{\tilde{\rho}}{\gamma}=\frac{\mathcal{K}_{\rho}}{\tilde{\zeta}(\tilde{\rho}^\text{r},\tilde{q}_0,\theta)^2}\tilde{\tau}_\text{Y}\frac{\tilde{q}(\tilde{\tau}_\text{Y},\tilde{\rho}^\text{r},\theta)}{\tilde{q}_0}\Bigl(1-\frac{\tilde{\rho}}{\tilde{\rho}^\text{s}(\tilde{\chi})}\Bigr),\notag\\
&\tilde{\tau}-\tilde{\tau}_\text{B}-\tilde{\tau}_\text{Y}=0. \notag
\end{align}
This system of PDEs comprises four equations in which both spatial and temporal derivatives occur. In order to achieve a numerically accurate solution, the original overall problem is parceled out into a large number of more easily solvable ODEs. With $\tilde{h}=h/b$ being the dimensionless height of the body, the interval $0<\tilde{x}_2<\tilde{h}$ is first decomposed into $n$ subintervals of the length $\Delta \tilde{h}=\tilde{h}/n$. The first and second spatial derivatives of the plastic slip $\beta$ can then be calculated using the finite difference approximations
\begin{equation}
\pdv{\beta}{\tilde{x}_2}=\frac{\beta_{i+1}-\beta_{i-1}}{2\Delta\tilde{h}},\quad\pdv[2]{\beta}{{\tilde{x}_2}}=\frac{\beta_{i+1}-2\beta_i+\beta_{i-1}}{{\Delta\tilde{h}}^2},
\label{finite_difference}
\end{equation}
where $\beta_i=\beta(i \, \Delta \tilde{h})$. The mean value of $\beta$, entering the equation for $\tau$, is calculated using the trapezoidal rule
\begin{displaymath}
\langle\beta\rangle =\frac{1}{n}\sum_{i=1}^{n-1} \beta_i .
\end{displaymath}
Let us express the dimensionless back stress from \eqref{backstress} in terms of these dimensionless derivatives:
\begin{equation}
%\label{dimensionless_backstress}
\tilde{\tau}_\text{B}=-\frac{k_1}{\bigl(1-k_2\vert\beta_{,\tilde{2}}\vert\bigr)^2}\beta_{,\tilde{2}\tilde{2}},
\end{equation}
where $\beta_{,\tilde{2}}$ and $\beta_{,\tilde{2}\tilde{2}}$ are computed in accordance with \eqref{finite_difference}, while
\begin{equation}
%\label{dimensionless_parameters}
k_1=\frac{\gamma_D}{\mu b^2} k \sin^2\varphi ,\quad k_2=k \vert\sin\varphi\vert, \quad k=\frac{1}{\rho^\text{s}b^2}.
\end{equation}
This back stress enters equation \eqref{dimensionless_PDE}$_4$, making it a coupled system of $n$ algebraic equations. Altogether, this procedure leads to a system of $4n$ ordinary differential-algebraic equations (DAE), which only have first derivatives with respect to $\gamma$. In the present study, a spatial discretization of the interval $(0,\tilde{h})$ into $n=1000$ subintervals as well as a temporal decomposition with a step size of $\Delta\gamma=10^{-6}$ is applied, whereby the latter is to be interpreted as a numerical shear increment. Finally, the usual DAE-system is solved with the internal Matlab subroutine \textit{ode15s}.

\begin{figure}[htb]
\centering
\includegraphics[width=0.6\textwidth]{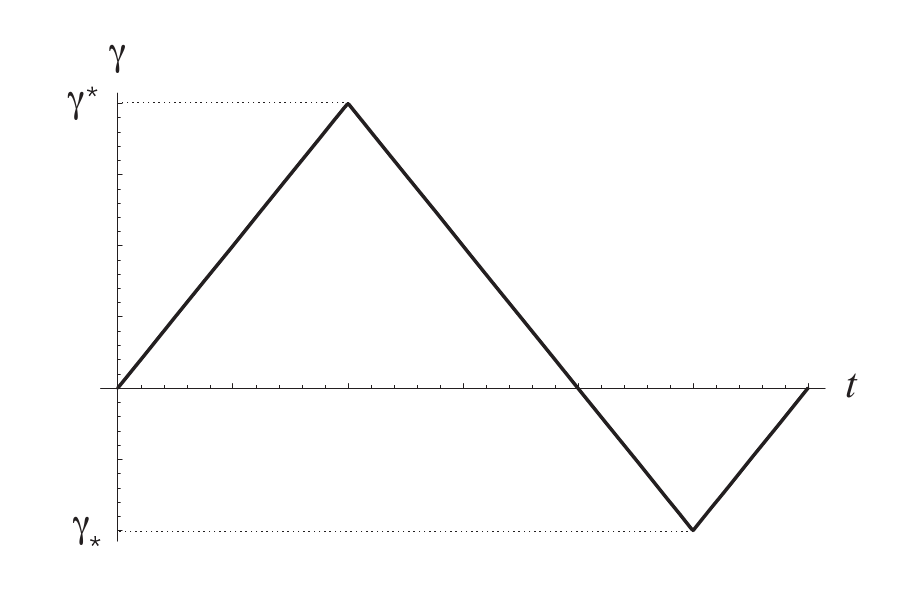}
\caption{A loading path.}
\label{fig:2}
\end{figure}

In order to examine the material behavior under load reversals, an entire loading path is simulated. This starts with the unloaded initial state, leads first to the maximum value $\gamma^\star$, then to the minimum value $\gamma_\star$ and finally again to complete unloading (see Fig.~\ref{fig:2}). The shear rate remains constant independent of the load direction. The load reversal scenario finds its realization in the reversal of equation \eqref{dimensionless_plastic_sliprate}$_2$ for the dimensionless plastic slip rate according to
\begin{equation}
\label{dimensionless_plastic_sliprate1}
\bar{q}(\tilde{\tau}_\text{Y},\tilde{\rho}^\text{r},\theta)=\sqrt{\tilde{\rho}^\text{r}}[\tilde{f}_\text{P}(-\tilde{\tau}_\text{Y},\tilde{\rho}^\text{r},\theta)-\tilde{f}_\text{P}(\tilde{\tau}_\text{Y},\tilde{\rho}^\text{r},\theta)].
\end{equation}
With \eqref{dimensionless_plastic_sliprate} or \eqref{dimensionless_plastic_sliprate1} the solution is obtained for any load direction by integrating the equations \eqref{dimensionless_PDE} with the corresponding $\tilde{q}$ or $\bar{q}$. In addition to the fulfillment of \eqref{dimensionless_PDE}, the continuity requirements at the transition points of the sections must be met in order to ensure the physical consistency of the solution. Therefore, when the target value $\gamma^\star$ or $\gamma_\star$ is reached, the calculated end values of a section are taken as initial values for the following section. In this way, a calculation algorithm is available that allows the variation of load modalities, such as load direction or speed, as well as the simulation of numerous load cycles to a comprehensive degree.

After computing the unknowns $\tilde{\tau}_\text{Y}$, $\tilde{\chi}$, $\tilde{\rho}$ and $\beta$ further parameters can be quantified. The dimensionless mean Schmid stress and the dimensionless mean back stress can be computed in an identical way by the relations
\begin{equation}
\frac{\bar{\tau}}{\mu}=\frac{1}{\tilde{h}}\int_0^{\tilde{h}}\tilde{\tau}\dd{x_2},\qquad \frac{\bar{\tau}_\text{Y}}{\mu}=\frac{1}{\tilde{h}}\int_0^{\tilde{h}}\tilde{\tau}_\text{Y} \dd{x_2}.
%\label{mean_stresses}
\end{equation}
The total number of dislocations per unit width is given by
\begin{equation}
N=\int_0^{h}\rho\dd{x_2}=\frac{b}{a^2}\int_0^{\tilde{h}}\tilde{\rho}\dd{\tilde{x}_2},
%\label{totaldislocation_perwidth}
\end{equation}
and the number of non-redundant dislocations per unit width is calculated by
\begin{equation}
%\label{excessdislocations_perwidth}
N^\text{g}=2\int_0^{\frac{h}{2}}\rho^\text{g}\dd{x_2}=\frac{2}{b}\vert\sin\varphi\vert\int_0^{\frac{h}{2}}\vert\beta_{,2}\vert\dd{x_2}=\frac{2}{b}\vert\sin\varphi\vert\beta_\text{m},
\end{equation}
where $\beta_\text{m}=\beta(\tilde{h}/2)$. In view of the symmetrical distribution of the non-redundant  dislocations over the height, the number from the only half of the height up to the center of the slab $\tilde{h}/2$ is computed, which should be multiplied by 2 to get $N^\text{g}$. Finally, the difference between $N$ and $N^\text{g}$ gives the number of redundant dislocations per unit width according to
\begin{equation}
%\label{redundant}
N^\text{r}=N-N^\text{g}.
\end{equation}

\begin{table}[t]
\centering
\begin{tabular}{llllllllll}
%\hline
\multicolumn{10}{c}{Material parameters} \\
\multicolumn{1}{c|}{$a$}  & \multicolumn{1}{c|}{$b$}  & \multicolumn{1}{c|}{$T_\text{P}$}  & \multicolumn{1}{c|}{$r$}  & \multicolumn{1}{c|}{$\tilde{\chi}_0$}  & \multicolumn{1}{c|}{$\mathcal{K}_{\rho}$}  & \multicolumn{1}{c|}{$\mathcal{K}_{\chi}$} & \multicolumn{1}{c|}{$\gamma_\text{D}$}  & \multicolumn{1}{c|}{$\mu$} & \multicolumn{1}{c}{$\nu$}\\
\hline
\multicolumn{1}{c|}{$10b$} & \multicolumn{1}{c|}{$0.255\text{nm}$} & \multicolumn{1}{c|}{$40822\text{K}$} & \multicolumn{1}{c|}{$0.0323$}  & \multicolumn{1}{c|}{$0.25$}  & \multicolumn{1}{c|}{$350$}   & \multicolumn{1}{c|}{$96.1$} & \multicolumn{1}{c|}{$\mu b^2$} & \multicolumn{1}{c|}{$50\text{GPa}$}  & \multicolumn{1}{c}{$0.33$}\\ 
\end{tabular}
%\end{table}
%\vspace{-0.33cm}
%\begin{table}[t]
\centering
\begin{tabular}{lllll}
%\hline
\multicolumn{5}{c}{Loading conditions} \\
\multicolumn{1}{c|}{$\gamma^{\star}$}  & \multicolumn{1}{c|}{$\gamma_{\star}$}  & \multicolumn{1}{c|}{$\tilde{q}_0$} & \multicolumn{1}{c|}{$\tilde{t}_0$} & \multicolumn{1}{c}{$T$} \\
\hline
\multicolumn{1}{c|}{$0.16$} & \multicolumn{1}{c|}{$-0.01$} & \multicolumn{1}{c|}{$10^{-12}$} & \multicolumn{1}{c|}{$10^{-12}\text{s}$} & \multicolumn{1}{c}{$298\text{K}$}   \\
\end{tabular}
%\end{table}
%\vspace{-0.33cm}
%\begin{table}[t]
\centering
\begin{tabular}{llll}
%\hline
\multicolumn{4}{c}{Initial data} \\
\multicolumn{1}{c|}{$\tilde{\tau}_\text{Y}(0)$}  & \multicolumn{1}{c|}{$\tilde{\rho}(0)$}  & \multicolumn{1}{c|}{$\tilde{\chi}(0)$}  & \multicolumn{1}{c}{$\beta(0)$}\\
\hline
\multicolumn{1}{c|}{$0$} & \multicolumn{1}{c|}{$6.25\times 10^{-5}$} & \multicolumn{1}{c|}{$0.18$} & \multicolumn{1}{c}{$0$}\\
\end{tabular}
\caption{Set of parameters}
\label{tab_Parameters}
\end{table}

Table~\ref{tab_Parameters} contains the set of parameters used in the numerical simulations. The list includes the parameters characterizing the material model, the loading conditions and the initial values. These data are consistent with those for copper at room temperature  \citep{langer2010thermodynamic,le2018thermodynamic}. From this Table we see that $k=(a^2/b^2)\exp (1/\tilde{\chi}_0)=5.46\times 10^3$. Note that copper is comparatively often the object of investigation in the numerical implementation of TDT, which is explained by its high thermal conductivity: The fast rate at which heat flows to the surrounding thermal bath during plastic deformations ensures an almost isothermal deformation process, so that the constant temperature assumed in theory finds its physical justification. The final shear of the opposing load $\gamma_\star$ is specifically defined in such a way that the specimen is stress-free after unloading. In addition, the dimensionless initial dislocation density represents an actual density of $\rho(0)=\tilde{\rho}(0)/(10b)^2\approx10^{13}\text{m}^{-2}$, which corresponds to a value typically found in metallic undeformed materials.

\section{Results of simulations} \label{S:4}
\subsection{Stress-strain curves} \label{S:4.1}

%%%%%%%%%%%%% FIGURE 3 %%%%%%%%%%%%
\begin{figure}[htb]
	\centering
	\includegraphics[width=.6\textwidth]{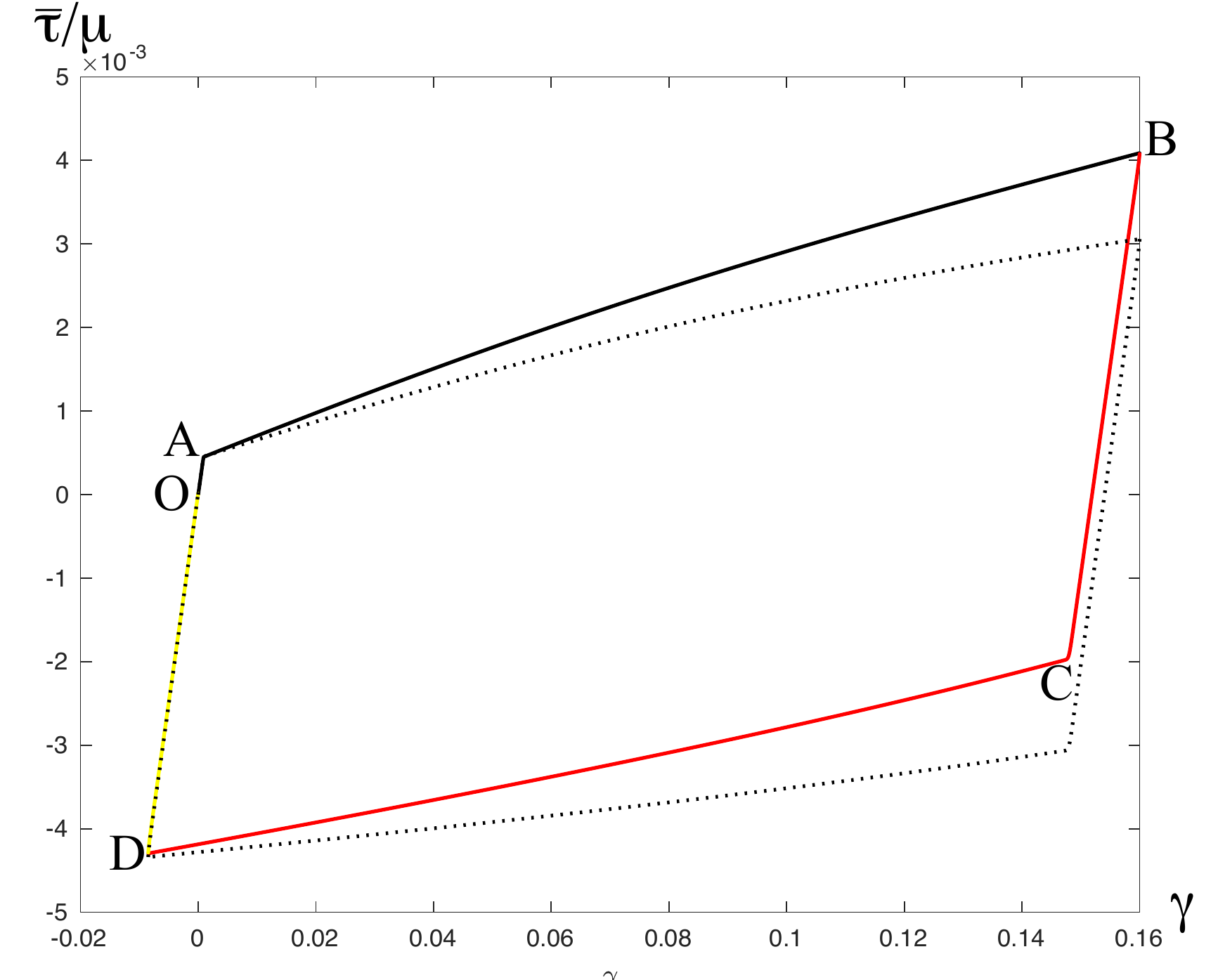}
	\caption{(Color online) Stress-strain curves at the strain rate $\tilde{q}_0=10^{-12}$ and room temperature, with $\tilde{h}=20000$ and $\varphi = 30^\circ$: (i) loading path OAB (black), (ii) load reversal BCD (red/dark gray), (iii) second load reversal DO (yellow/light gray), (iv) flow stress versus strain (dotted black curve).}
	\label{fig:3}
\end{figure}
%%%%%%%%%%%%%%%%%%%%%%%%%%%%%%%%%%%%% 

Fig.~\ref{fig:3} presents the rescaled averaged Schmid stress, $\bar{\tau}/\mu$ (bold line), and rescaled averaged flow stress, $\bar{\tau}_\text{Y}/\mu$ (dotted line), versus the shear strain $\gamma$ over the complete loading path shown in Fig.~\ref{fig:2}. The dimensionless height of the slab $\tilde{h}=20000$ and the angle $\varphi = 30^\circ$ are chosen. The plots of the two averaged stresses versus the shear strain $\gamma$ curves show identical behavior: Starting with the loading phase from the origin O, both initially develop in an identical manner along the elastic region on the line OA, before moving on from the identical initial yielding point A into the plastic region AB exhibiting the work hardening as $\gamma$ goes further to $\gamma^*$. Remarkable for the plastic region are the different slops of the two curves (hardening rates). Note that, with increasing shear strain the isotropic hardening due to $\bar{\tau}_\text{Y}$ decreases, while the kinematic hardening due to the back stress $\bar{\tau}_\text{B}=\bar{\tau }-\bar{\tau}_\text{Y}$ increases. This first loading phase is followed by the load reversal phase in which $\gamma$ decreases from $\gamma^*$ to $\gamma_*$. The stress-strain curve also begins with the elastic line BC. Note that the elastic line of $\bar{\tau}_\text{Y}/\mu$ is parallel to, but differs from that of $\bar{\tau}/\mu$ at this stage. The yielding transition occurs at C, where the magnitude of the stress is much lower than that at the end of the loading path exhibiting the Bauschinger effect which will be explained later. Then the stress-strain curve follows the plastic region on the line CD which shows the increasing hardening as $\gamma$ decreases to $\gamma_*$. The second load reversal phase, in which $\gamma$ rises from $\gamma_*$ to zero, again shows the elastic behavior on the line DO.

%%%%%%%%%%%%% FIGURE 4 %%%%%%%%%%%%%%%%%%%
\begin{figure}[htb]
	\centering
	\includegraphics[width=.6\textwidth]{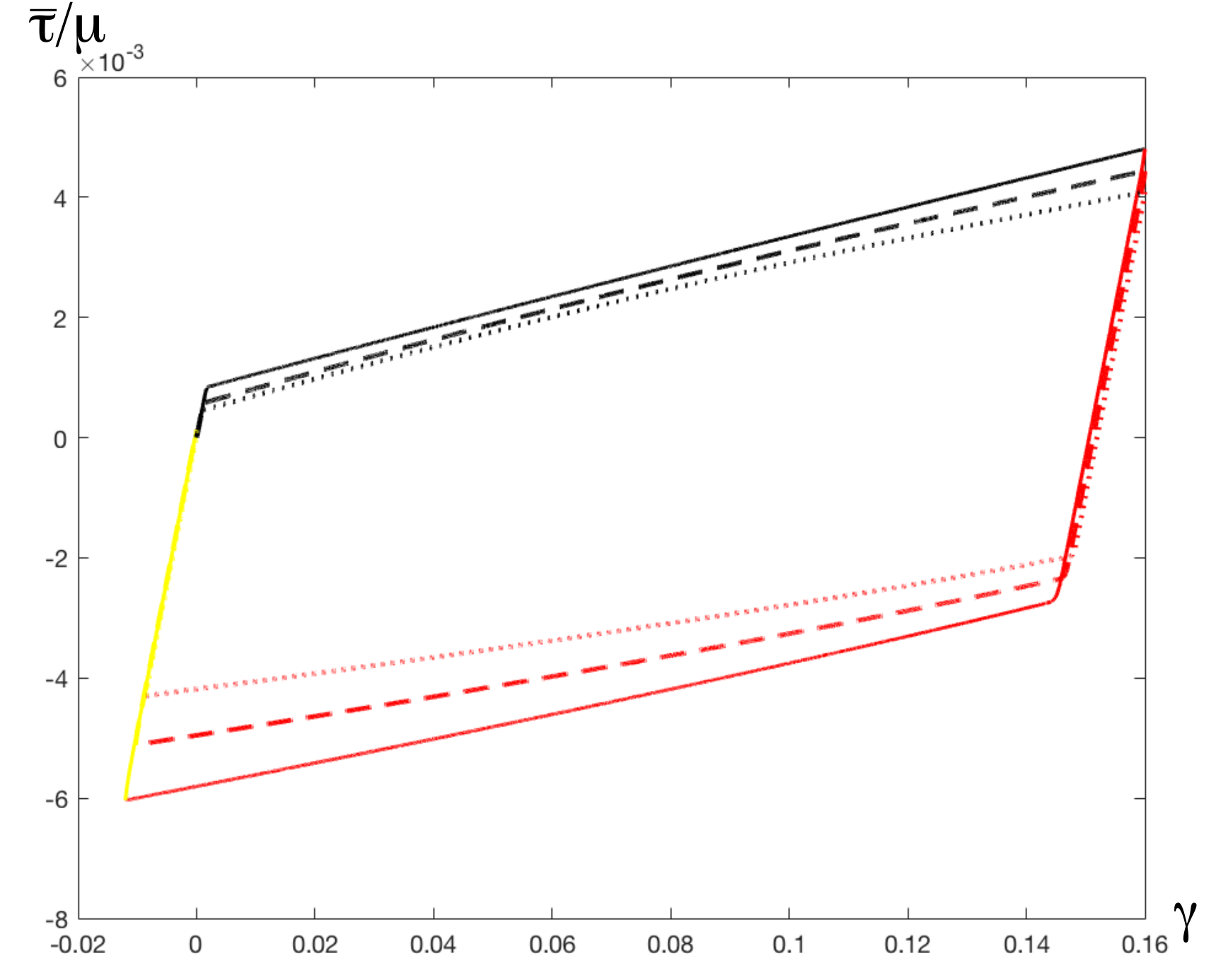}
	\caption{(Color online) Stress-strain curves at three strain rates $\tilde{q}_{01}=10^{-4}$ (solid lines), $\tilde{q}_{02}=10^{-8}$ (dashed lines) and $\tilde{q}_{03}=10^{-12}$ (dotted lines) and at room temperature, with $\tilde{h}=20000$ and $\varphi = 30^\circ$.}
	\label{fig:4}
\end{figure}
%%%%%%%%%%%%%%%%%%%%%%%%%%%%%%%%%%%%% 

The stress-strain curves are sensitive with respect to the shear rate. In order to show this we plot in Fig.~\ref{fig:4} the rescaled averaged Schmid stress, $\bar{\tau}/\mu$ versus the shear strain $\gamma$ over the complete loading path for three different shear rates, where the solid lines corresponds to the rate $\tilde{q}_{01}=10^{-4}$, the dashed lines $\tilde{q}_{02}=10^{-8}$ and the dotted lines $\tilde{q}_{03}=10^{-12}$. The resulting curves in Fig.~\ref{fig:4} confirm the findings of the rate-dependent study, according to which faster strain rates imply larger Schmid stresses. Moreover, this Figure allows a statement to be made regarding the sensitivity of isotropic and kinematic hardening to a variation in shear rate: The rate dependence of the work hardening is mainly due to the isotropic hardening, whereas the kinematic hardening proves to be relatively less sensitive to a variation of the shear rate. 

%%%%%%%%%%%%% FIGURE 5 %%%%%%%%%%%%%%%%%%%
\begin{figure}[htb]
	\centering
	\includegraphics[width=.6\textwidth]{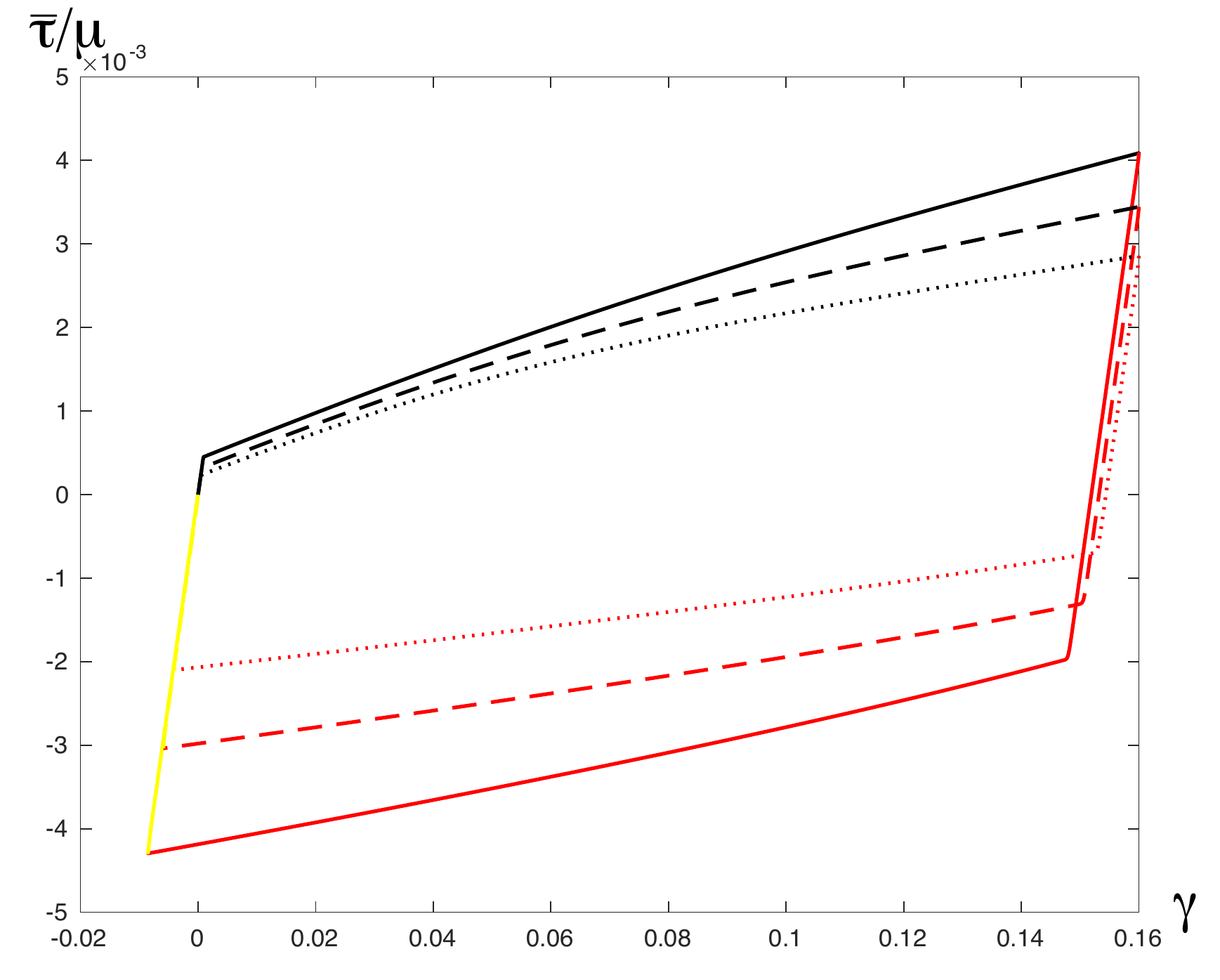}
	\caption{(Color online) Stress-strain curves at the strain rate $\tilde{q}_{01}=10^{-12}$ and at three different (ordinary) temperatures $T_1=298\text{K}$ (solid lines), $T_2=498\text{K}$ (dashed lines) and $T_3=698\text{K}$ (dotted lines), with $\tilde{h}=20000$ and $\varphi = 30^\circ$.}
	\label{fig:5}
\end{figure}
%%%%%%%%%%%%%%%%%%%%%%%%%%%%%%%%%%%%% 

We also study the sensitivity of the stress-strain curves $\bar{\tau}(\gamma )/\mu$ with respect to the ordinary temperature of the surrounding thermal bath. For a more detailed evaluation of the work hardening during the plastic deformation, the rescaled averaged Schmid stress versus $\gamma$ for the three ordinary  temperatures of the thermal bath $T_1=298\text{K}$ (solid lines), $T_2=498\text{K}$ (dashed lines) and $T_3=698\text{K}$ (dotted lines) accompanying a complete load cycle are plotted in Fig.~\ref{fig:5}. This Figure illustrates the physically reasonable tendency that, with increasing $T$, the Schmid stress together with the work hardening decreases. The physical explanation is simple: The increase in temperature facilitates the "triggering" of dislocations from the immobile to the free state, which leads to an increase in the dislocation depinning rate and together with it the plastic strain rate. Consequently, an increased $T$ is reflected in the reduction of the work hardening. This explanation also justifies the shortening of the elastic ranges as the initial yielding is reached faster with increasing temperature. Similar to the strain rate sensitivity, the temperature sensitivity mainly concerns the isotropic hardening and to a much lesser extent the kinematic hardening.

%%%%%%%%%%%%% FIGURE 6 %%%%%%%%%%%%%%%%%%%
\begin{figure}[htb]
	\centering
	\includegraphics[width=.6\textwidth]{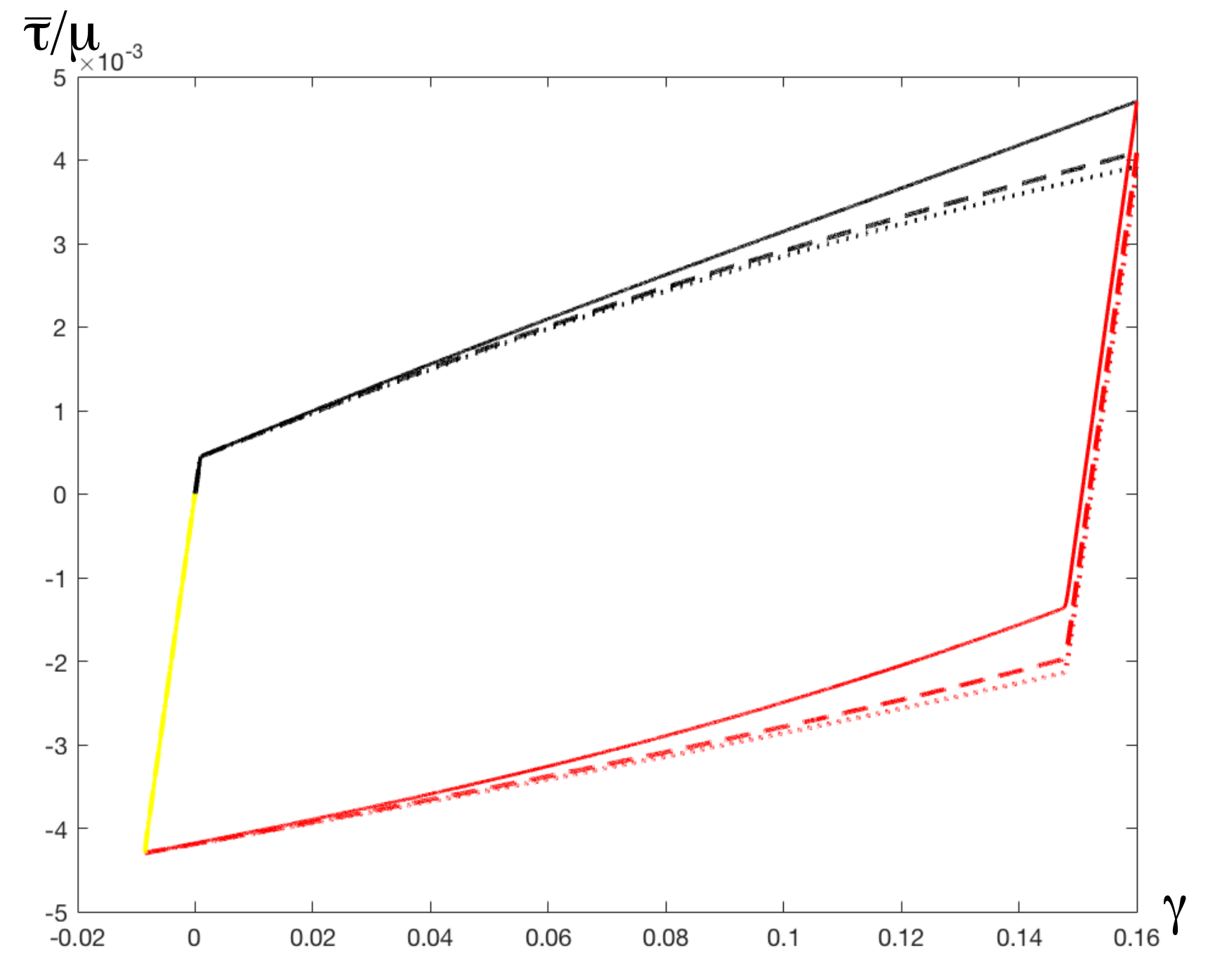}
	\caption{(Color online) Stress-strain curves at the strain rate $\tilde{q}_{01}=10^{-12}$ and room temperature, for three different sample heights $\tilde{h}_1=10000$ (bold lines), $\tilde{h}_2=20000$ (dashed lines) and $\tilde{h}_3=100000$ (dotted lines), with $\varphi = 30^\circ$.}
	\label{fig:6}
\end{figure}
%%%%%%%%%%%%%%%%%%%%%%%%%%%%%%%%%%%%% 

The next factor of interest is the height $h$ of the sheared slab. To analyze the size effect the stress-strain curves $\bar{\tau}(\gamma )/\mu$ over the complete loading path are plotted in Fig.~\ref{fig:6} using three dimensionless slab's heights $\tilde{h}_1=10000$ (bold lines), $\tilde{h}_2=20000$ (dashed lines) and $\tilde{h}_3=100000$ (dotted lines). The simulation results from this Figure show the tendency that the decrease of the sample height causes the increase of $\bar{\tau}$, an observation which can be summarized in the relation $\bar{\tau}(\tilde{h}_1)>\bar{\tau}(\tilde{h}_2)>\bar{\tau}(\tilde{h}_3)$ with $\tilde{h}_1<\tilde{h}_2<\tilde{h}_3$. Besides, the hardening rate in the plastic region increases with the reduction in the sample height (``smaller is stronger''). The more detailed analysis of the flow stress $\bar{\tau}_\text{Y}$ shows that this size effect is solely due to the kinematic hardening and the pile-up of non-redundant dislocations. For instance, despite the increase of $\bar{\tau}$ after the end of the loading phase, the spans of all three flow areas turn out to be almost invariant, which demonstrates the size independence of the isotropic hardening. A further qualitative confirmation of this claim is the common intersection of all three stress-strain curves at the onset of plastic yielding and after the load reversal. For heights greater than $100000b$, which correspond to almost macroscopic specimens, the size effect is less pronounced and only becomes apparent at greater strains.

%%%%%%%%%%%%% FIGURE 7 %%%%%%%%%%%%%%%%%%%
\begin{figure}[htb]
	\centering
	\includegraphics[width=.6\textwidth]{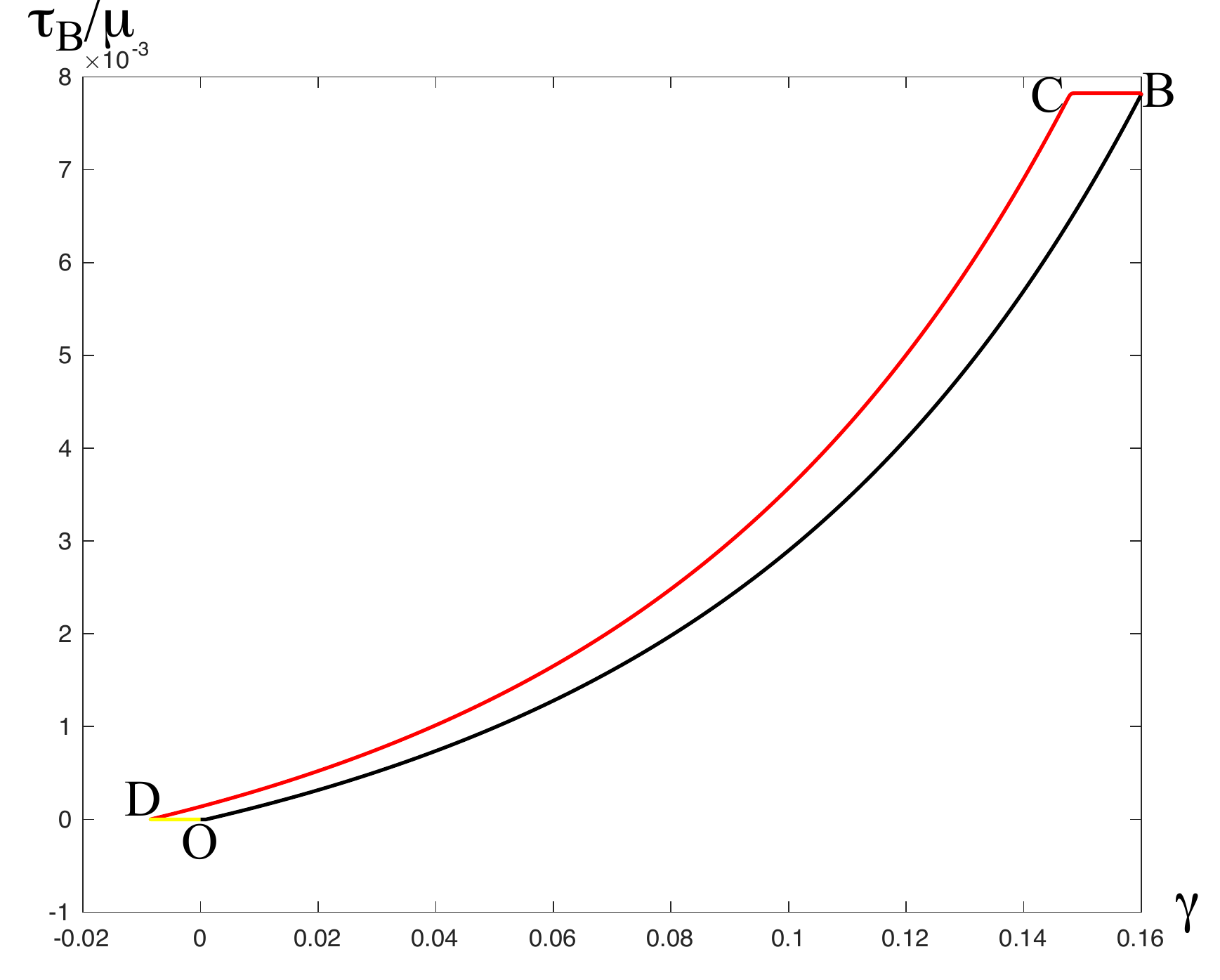}
	\caption{(Color online) Normalized back stress $\tau_\text{B}/\mu$ near the boundary versus $\gamma$ at the strain rate $\tilde{q}_{01}=10^{-12}$ and room temperature, with $\tilde{h}=20000$ and $\varphi = 30^\circ$: (i) loading path (black), (ii) load reversal (red/dark gray), and (iii) second load reversal (yellow/light gray).}
	\label{fig:7}
\end{figure}
%%%%%%%%%%%%%%%%%%%%%%%%%%%%%%%%%%%%% 

To explain the Bauschinger effect we now show the evolution of the normalized back stress $\tau_\text{B}/\mu$ computed in the thin boundary layer as function of $\gamma$ over the complete loading path in Fig.~\ref{fig:7}. After the starting point O and the subsequent relatively short elastic line OA without the back stress, $\tau_\text{B}/\mu$ increases with the applied shear $\gamma$ over the entire loading phase ending in the point B. The load reversal phase is also divided into a constant elastic line BC with constant $\tau_\text{B}/\mu$ and a falling plastic line CD. The positive back stress at C causes the lower magnitude of the stress required for initiating the second yielding than that at the first yielding point A. The plastic line ends with the reaching of the zero value for $\tau_\text{B}/\mu$ in point D, from which the unloading phase follows, until the initial state without applied shear is reached again in point O. Finally, the parameter study on the back stress (see sub-section 4.5) is used to validate the hypotheses expressed in the investigations of the rate, temperature and sample size dependence for $\bar{\tau}_\text{Y}$ and $\bar{\tau}$ with respect to the hardening behavior. According to this, a comparatively constant development of the back stress with respect to the variation of shear rate and temperature witnesses the insensitivity of kinematic hardening to these parameters. On the other hand, the change in sample size affects the back stress in the boundary layer in such a way that $\tau_\text{B}$ decreases with increasing slab's height.

\subsection{Distributions and evolution of plastic slip} \label{S:4.2}

%%%%%%%%%%%%% FIGURE 8 %%%%%%%%%%%%%%%%%%%
\begin{figure}[htb]
	\centering
	\includegraphics[width=.6\textwidth]{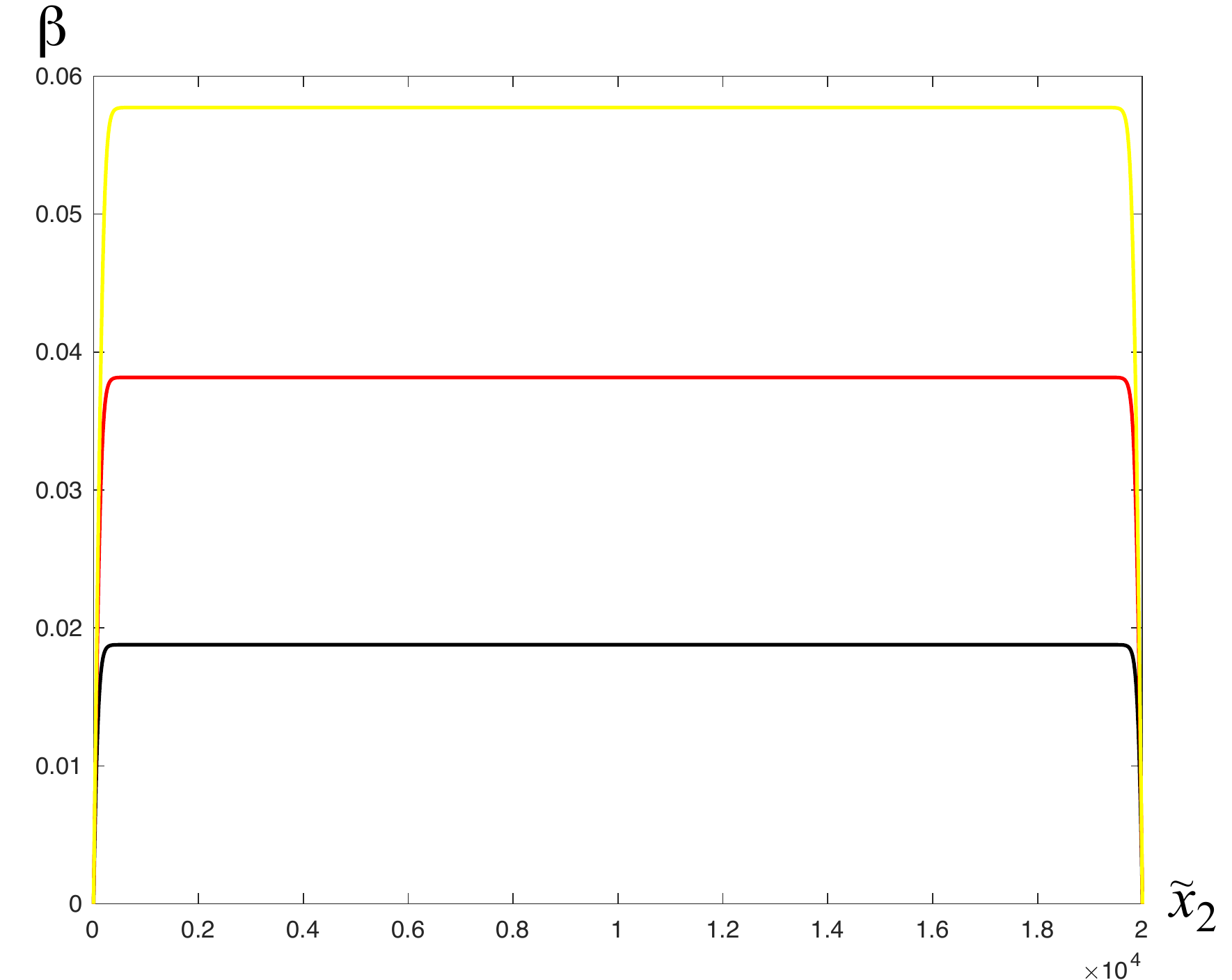}
	\caption{(Color online) Evolution of $\beta(\tilde{x}_2)$ at the strain rate $\tilde{q}_0=10^{-12}$ and room temperature, with $\tilde{h}=20000$ and $\varphi = 30^\circ$, during the loading along AB: (i) $\gamma =0.04$ (black), (ii) $\gamma =0.08$ (red/dark gray), (iii) $\gamma =0.12$ (yellow/light gray).}
	\label{fig:8}
\end{figure}
%%%%%%%%%%%%%%%%%%%%%%%%%%%%%%%%%%%%% 

Fig.~\ref{fig:8} represents the plastic slip $\beta(\tilde{x}_2,\gamma)$ during the loading phase for the three shear amounts $\gamma_1=0.04$, $\gamma_2=0.08$ and $\gamma_3=0.12$, simulated with the use of the standard parameter set from Table~\ref{tab_Parameters}. The plot results in symmetrical, plateau-shaped curves, each of which reaches its plateau value starting and ending with the zero value within a comparatively short length of boundary layers. The solution also shows the basic relative increase in plastic slip during increasing shear stress and agrees qualitatively with the approximate solution found in \citep{le2018athermodynamic}. The quantitative comparison is difficult due to the different choices for the energy of non-redundant dislocations. On the background of the physical interpretation of $\beta(\tilde{x}_2,\gamma)$, the result appears to make sense. The non-uniformity of the plastic slip causes non-redundant dislocations whose Burgers vectors do not cancel each other out. The relationship here is that the absence of non-redundant dislocations is characterized by the zero slope of the curve in the middle, whereas the positive and negative slopes near the boundaries indicate that non-redundant dislocations of opposite signs pile up against the lower and upper grain boundaries. The plastically deformed specimen thus has two thin boundary layers on the top and bottom in which the non-redundant (geometrically necessary) dislocations accumulate, whereas the predominant, central area does not contain this type of dislocations. 

%%%%%%%%%%%%% FIGURE 9 %%%%%%%%%%%%%%%%%%%
\begin{figure}[htb]
	\centering
	\includegraphics[width=.6\textwidth]{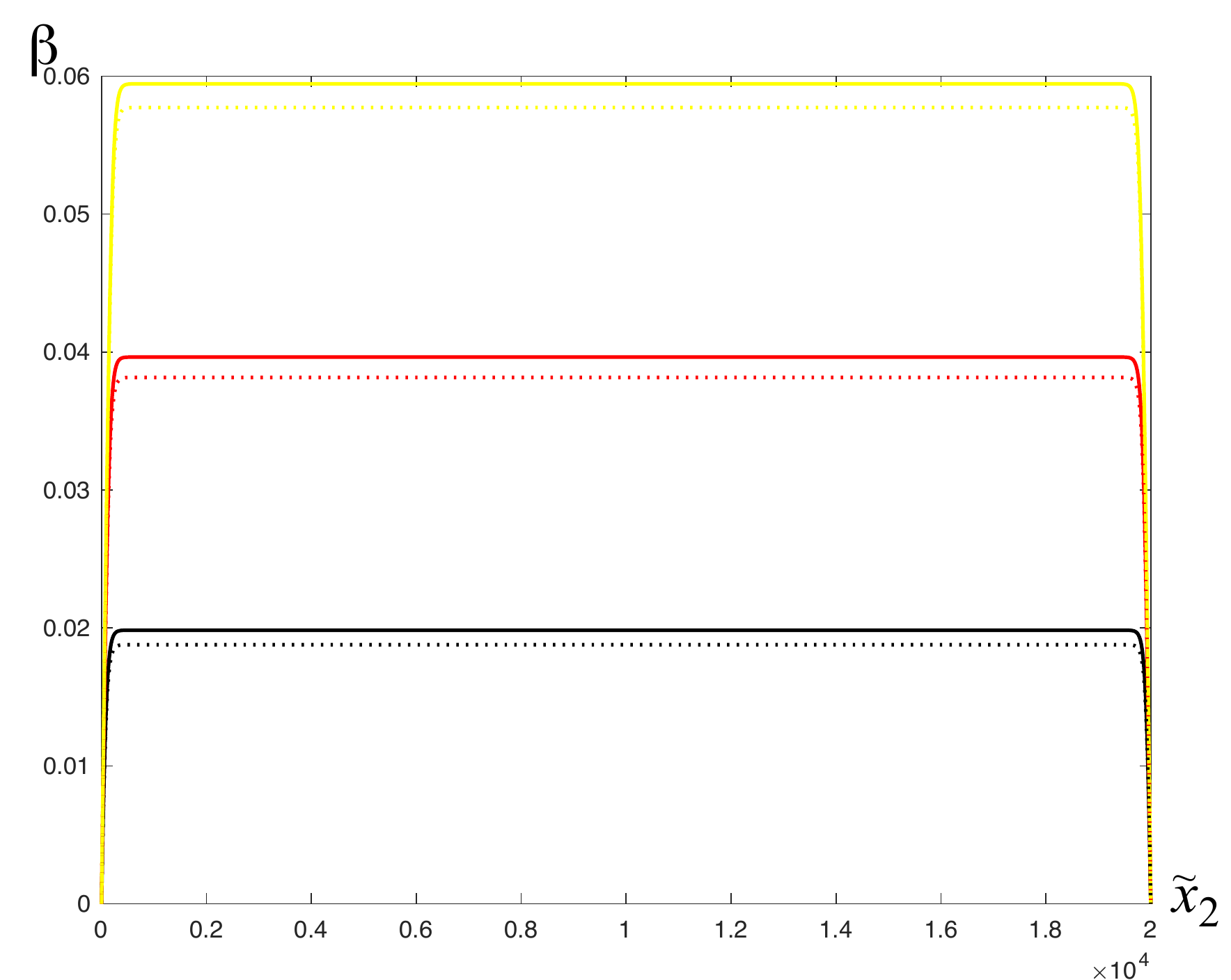}
	\caption{(Color online) Evolution of $\beta(\tilde{x}_2)$ at the strain rate $\tilde{q}_0=10^{-12}$ and room temperature, with $\tilde{h}=20000$ and $\varphi = 30^\circ$, during the loading (dotted lines) and load reversal (bold lines): (i) $\gamma =0.04$ (black), (ii) $\gamma =0.08$ (red/dark gray), (iii) $\gamma =0.12$ (yellow/light gray).}
	\label{fig:9}
\end{figure}
%%%%%%%%%%%%%%%%%%%%%%%%%%%%%%%%%%%%% 

Fig.~\ref{fig:9} shows the evolution of $\beta(\tilde{x}_2,\gamma)$ during the load reversal phase, and, for comparison purpose, also $\beta(\tilde{x}_2,\gamma)$ during the loading phase. Again, the data from Table~\ref{tab_Parameters}, $\tilde{h}=20000$ and the angle $\varphi = 30^\circ$ are used for the simulation. The load reversal phase is characterized by basically identical distribution of the plastic slip as those of the load phase: In addition to the steep slopes near the boundaries, there is a plateau in the middle section where non-redundant dislocations are not present. The values in the plateau have increased moderately compared to those during the loading phase. The results show a physically reasonable behavior: Despite the change in load direction, the distribution of non-redundant dislocations is maintained so that the $\beta$ profile is present in an analogous manner. The almost constant difference is explained by the fact that from the onset of plastic flow, regardless of the direction of loading, non-redundant dislocations are formed at the same rate. 

\subsection{Evolution of dislocations} \label{S:4.3}

%%%%%%%%%%%%% FIGURE 10 %%%%%%%%%%%%%%%%%%%
\begin{figure}[htb]
	\centering
	\includegraphics[width=.6\textwidth]{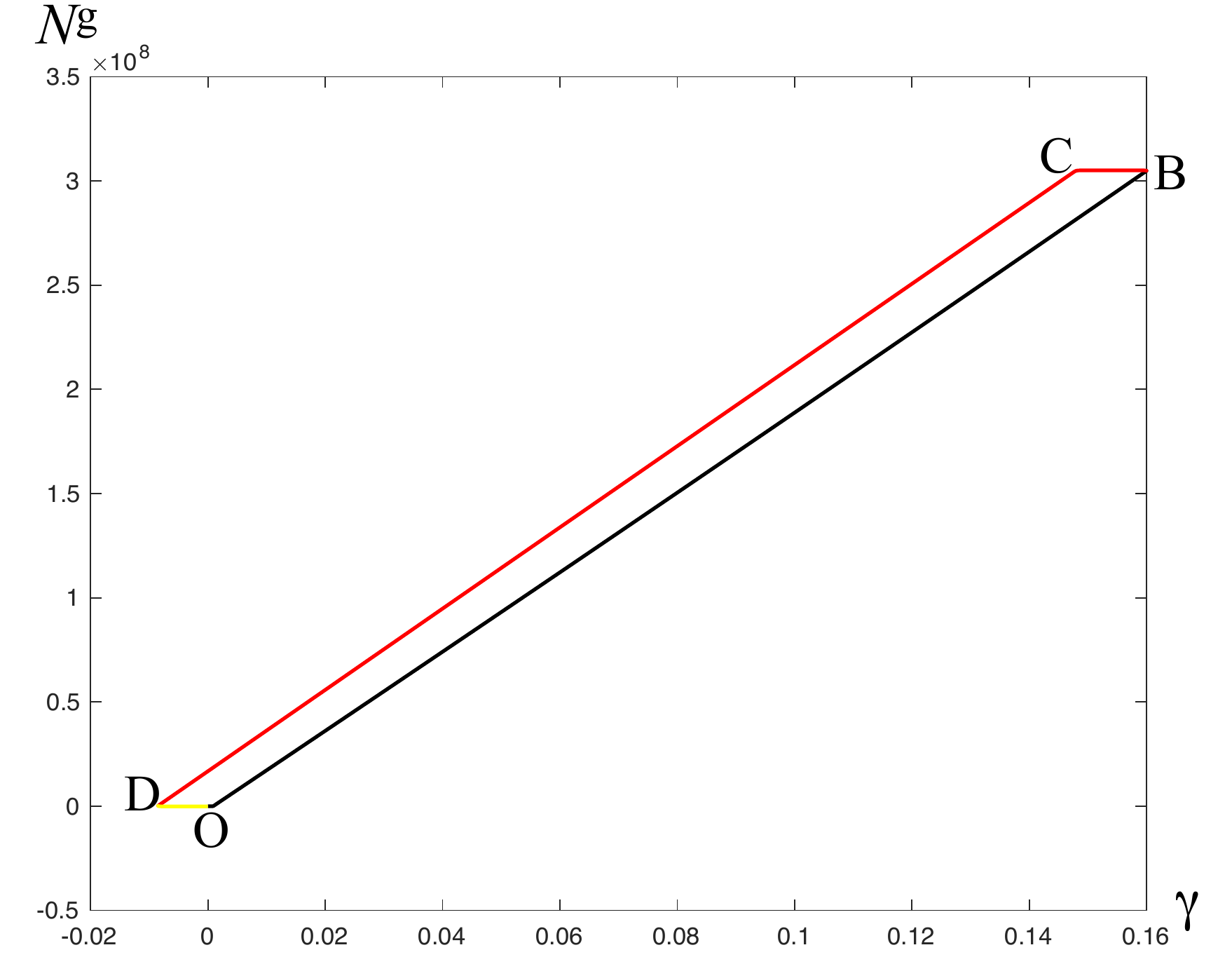}
	\caption{(Color online) Number of non-redundant dislocations per unit width $N^\text{g}$ versus $\gamma$ at the strain rate $\tilde{q}_0=10^{-12}$ and room temperature, with $\tilde{h}=20000$ and $\varphi = 30^\circ$: (i) loading path (black), (ii) load reversal (red/dark gray), (iii) second load reversal (yellow/light gray).}
	\label{fig:10}
\end{figure}
%%%%%%%%%%%%%%%%%%%%%%%%%%%%%%%%%%%%% 

Fig.~\ref{fig:10} shows the evolution of the number of non-redundant dislocations (per unit width) $N^\text{g}(\gamma)$ during the complete load cycle which is qualitatively identical to that of the back stress $\tau_\text{B}(\gamma)$ in Fig.~\ref{fig:7}. At first these show a rising, then a falling tendency, whereby the rising line $\text{AB}$ and falling line $\text{CD}$ are separated by the two horizontal lines $\text{BC}$ and $\text{AD}$. While initially no non-redundant dislocations are present in the crystal, their maximum number, when the maximum shear $\gamma=0.16$ is reached, is about $3.1\times 10^{8}\text{m}^{-1}$. To obtain the dislocation density $\rho^\text{g}$, we must divide this number by the height of the slab, resulting in $6.1\times 10^{12}\text{m}^{-2}$. This number $N^\text{g}$ changes only during the plastic deformation, whereas it remains constant (frozen) during the elastic deformation. The decrease of $N^\text{g}$ during the load reversal can be explained as follows: The presence of the positive back stress reduces the magnitude of shear stress required for pulling the non-redundant dislocations back to the center of the specimen. There, the non-redundant dislocations of opposite signs meet and annihilate each other, so the number of non-redundant dislocations reduces gradually to zero along the curve CD. 

%%%%%%%%%%%%% FIGURE 11 %%%%%%%%%%%%%%%%%%%
\begin{figure}[htb]
	\centering
	\includegraphics[width=.6\textwidth]{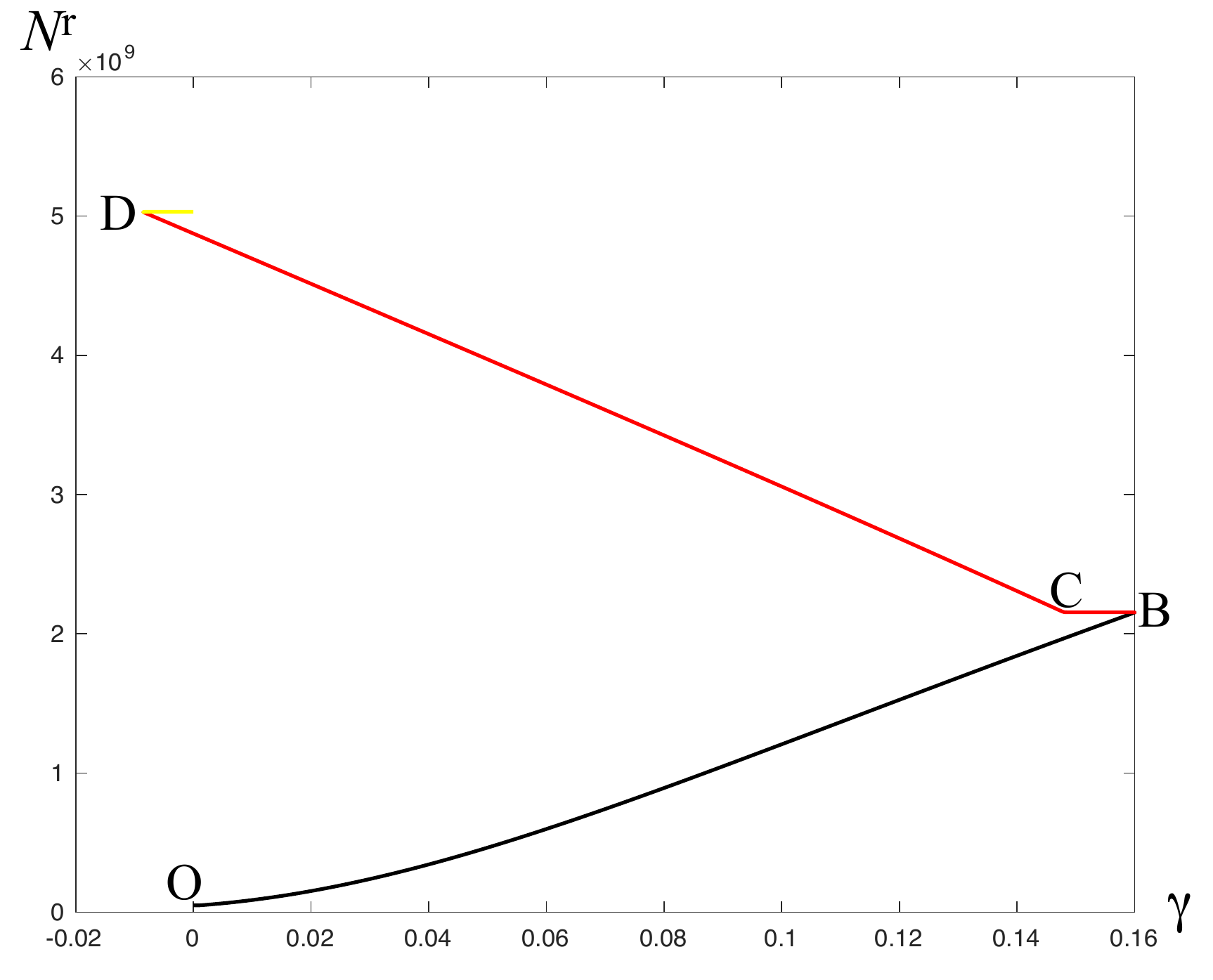}
	\caption{(Color online) Number of redundant dislocations per unit width $N^\text{r}$ versus $\gamma$ at the strain rate $\tilde{q}_0=10^{-12}$ and room temperature, with $\tilde{h}=20000$ and $\varphi = 30^\circ$: (i) loading path (black), (ii) load reversal (red/dark gray), (iii) second load reversal (yellow/light gray).}
	\label{fig:11}
\end{figure}
%%%%%%%%%%%%%%%%%%%%%%%%%%%%%%%%%%%%% 

Fig.~\ref{fig:11} shows the essentially different evolution of the number of redundant dislocations (per unit width) $N^\text{r}(\gamma)$ as compared to that of $N^\text{g}(\gamma)$. The most remarkable difference between the behaviors of $N^\text{r}$ and $N^\text{g}$ is that the former increases further along CD during the load reversal while the latter decreases to zero. Thus, along CD the material is closer to the steady state than along AB, and consequently, the hardening rate of the stress-strain curve of CD shown in Fig.~\ref{fig:3} must be less than that of AB. This asymmetry between loadings in opposite directions becomes more pronounced as $\gamma^*$ increases. Note that the total number of dislocations $N(\gamma)$ behaves in exactly the same way as $N^\text{r}(\gamma)$. The numerical difference between them is due to the number of non-redundant dislocations, which, for small strains, is still much smaller than $N^\text{r}(\gamma)$. 

\subsection{Evolution of configurational temperature} \label{S:4.4}

%%%%%%%%%%%%% FIGURE 12 %%%%%%%%%%%%%%%%%%%
\begin{figure}[htb]
	\centering
	\includegraphics[width=.6\textwidth]{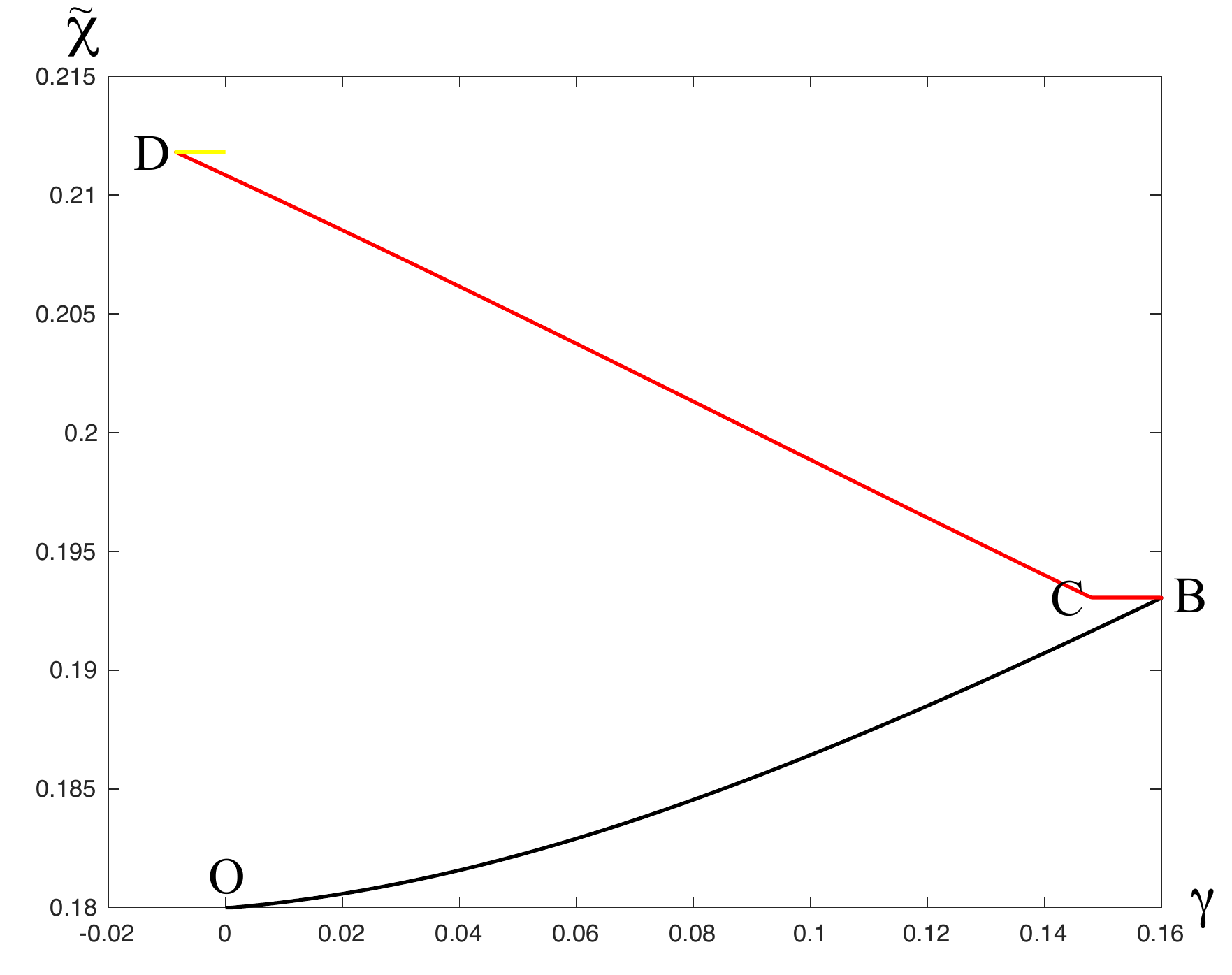}
	\caption{(Color online) The configurational temperature $\tilde{\chi}(\tilde{c}/2)$ versus $\gamma$ at the strain rate $\tilde{q}_0=10^{-12}$ and room temperature, with $\tilde{h}=20000$ and $\varphi = 30^\circ$: (i) loading path (black), (ii) load reversal (red/dark gray), (iii) second load reversal (yellow/light gray).}
	\label{fig:12}
\end{figure}
%%%%%%%%%%%%%%%%%%%%%%%%%%%%%%%%%%%%% 

The evolution of $\tilde{\chi}(\tilde{h}/2)$ versus $\gamma$ turns out to be similar to that of $N^\text{r}$ (or $N$) and is shown in Fig.~\ref{fig:12}. In addition to the small horizontal elastic lines, the plastic lines are characterized by the positive slops. Thus, after a short stagnation along BC the configurational temperature increases further during the load reversal, what moves the system closer to the steady state. Note that the correlation between the curves of the configuration temperature $\tilde{\chi}(\gamma)$ and the dislocation number $N(\gamma)$ has its root in the mathematically comparable category of the DE of these two quantities according to \eqref{dimensionless_PDE}$_2$ and \eqref{dimensionless_PDE}$_3$: Both right-hand sides are of limited decreasing character, whereby the latter additionally depends on the weakly varying function ${\nu}$, whose influence proves to be negligible in the present context. From a thermodynamic point of view, the increase in $\rho$ and $\chi$ causes the dislocation multiplication to decrease or, in other words, the dislocation annihilation to intensify. Thus the higher the disorder temperature, the faster the saturation effect affects the number of dislocations.

\subsection{Summary of the parameter study} \label{S:4.5}

We perform the detailed parameter study of all quantities in the previous subsections, the results of which are summarized in Table~\ref{tab_Result}. This table shows the qualitative changes of all variables during the increase of the four influencing parameters. The binary evaluation scale represents the relations exclusively qualitatively, where a ``$+$'' basically means the relative increase and a ``$-$'' the relative decrease. The ``$0$'' indicates an inert behavior towards parameter variation.

\begin{table}[h!]
\centering
\begin{tabular}{llllllllll}
%\hline
\multicolumn{1}{c|}{}  & \multicolumn{1}{c|}{$\bar{\tau}$}  & \multicolumn{1}{c|}{$\bar{\tau}_\text{Y}$}  & \multicolumn{1}{c|}{$\quad\tau_\text{B}$}  & \multicolumn{1}{c|}{$N^\text{r}$}  & \multicolumn{1}{c|}{$N^\text{g}$}  & \multicolumn{1}{c|}{$\tilde{\chi}$}  & \multicolumn{1}{c|}{$\beta$}  & \multicolumn{1}{c|}{kinem.}  & \multicolumn{1}{c}{isotr.}\\
\multicolumn{1}{c|}{}  &  \multicolumn{1}{c|}{} &  \multicolumn{1}{c|}{} &  \multicolumn{1}{c|}{} &  \multicolumn{1}{c|}{} &  \multicolumn{1}{c|}{} &  \multicolumn{1}{c|}{} &  \multicolumn{1}{c|}{} & \multicolumn{1}{c|}{hard.}  & \multicolumn{1}{c}{hard.}\\
\hline
\multicolumn{1}{c|}{$\tilde{q}_{0}\uparrow$} & \multicolumn{1}{c|}{$+++$} & \multicolumn{1}{c|}{$+++$} & \multicolumn{1}{c|}{$-$} & \multicolumn{1}{c|}{$--$} & \multicolumn{1}{c|}{$-$} & \multicolumn{1}{c|}{$++$} & \multicolumn{1}{c|}{$-$}  & \multicolumn{1}{c|}{$-$} & \multicolumn{1}{c}{$+++$} \\
\hline 
\multicolumn{1}{c|}{$T\uparrow$} & \multicolumn{1}{c|}{$---$} & \multicolumn{1}{c|}{$---$} & \multicolumn{1}{c|}{$+$} & \multicolumn{1}{c|}{$++$} & \multicolumn{1}{c|}{$+$} & \multicolumn{1}{c|}{$--$} & \multicolumn{1}{c|}{$+$}  & \multicolumn{1}{c|}{$+$} & \multicolumn{1}{c}{$---$} \\
\hline 
\multicolumn{1}{c|}{$\tilde{h}\uparrow$} & \multicolumn{1}{c|}{$---$} & \multicolumn{1}{c|}{$0$} & \multicolumn{1}{c|}{$---$} & \multicolumn{1}{c|}{$+++$} & \multicolumn{1}{c|}{$-$} & \multicolumn{1}{c|}{$0$} & \multicolumn{1}{c|}{$-$}  & \multicolumn{1}{c|}{$---$} & \multicolumn{1}{c}{$0$} \\
\hline 
\multicolumn{1}{c|}{$\varphi\uparrow$} & \multicolumn{1}{c|}{$--$} & \multicolumn{1}{c|}{$--$} & \multicolumn{1}{c|}{$++$} & \multicolumn{1}{c|}{$--$} & \multicolumn{1}{c|}{$++$} & \multicolumn{1}{c|}{$--$} & \multicolumn{1}{c|}{$--$}  & \multicolumn{1}{c|}{$++$} & \multicolumn{1}{c}{$--$}  \\
\end{tabular}
\caption{Effect of the increase of the influencing parameters on the results}
\label{tab_Result}
\end{table}

In particular, the list demonstrates the significant sensitivity of the Schmid and flow stresses, the number of redundant dislocation and the configurational temperature on the variation of strain rate and temperature. The variation in the loading modality, however, has a comparatively marginal effect on the back stress and the number of non-redundant dislocations. In contrary, the enlargement of the sample causes changes of completely different characteristics which mainly concern the non-redundant dislocations and the kinematic hardening. While the number of non-redundant dislocations does not change much, which leads to a reduction in the density of this type of dislocations due to the increase in height, the number of redundant dislocations increases simultaneously to the extent that the corresponding density remains constant and therefore no noticeable change in the flow stress can be observed. Another remarkable feature is the analogy between the Schmid stress and the back stress, both of which have the identical dependence on the change in height. The last variable, the angle of inclination $\varphi$ characterizing the orientation of the slip system, causes a manifold change of the quantities: On the one hand, this is expressed in the relative reduction in the number of redundant dislocations, the configurational temperature and consequently the Schmid and flow stress, while on the other hand, the number of non-redundant dislocations increases relatively, as does the back stress.

\section{Conclusion} \label{S:5}
The thermodynamic approach, which incorporates configurational temperature and non-redundant dislocations, is proving to be an effective tool for the construction of dislocation-based predictive plasticity. The system of PDEs derived from the TDT has been transformed into a system of DAEs by discretization and then solved numerically efficiently. Parameter studies prove the increase of the Schmid stress on the one hand by increasing the shear rate and on the other hand by reducing the ordinary temperature, the grain size and the angle of inclination of the slip direction. The back stress is primarily a dominating factor for the size effect, whereas the isotropic hardening is insensitive to the change in size. The physical explanation of the Bauschinger effect, which is based on back stress and the retraction and annihilation of non-redundant dislocations, is convincing. Based on this theory, the asymmetry of work hardening between loads in opposite directions could be explained and predicted.

The computer-aided implementation of the present study is characterized by a comparatively high numerical performance, which strengthens the integration of the basic concept in higher-dimensional problems based on finite element computation. The numerical realization presented in this paper can be used as a subroutine that quantifies the development of the microstructure after each loading step from the structural-mechanical data of the finite element calculation. 

\section{Conflict of interest statement} \label{S:6}
On behalf of all authors, the corresponding author states that there is no conflict of interest.


\begin{thebibliography}{38}
\providecommand{\natexlab}[1]{#1}
\providecommand{\url}[1]{{#1}}
\providecommand{\urlprefix}{URL }
\expandafter\ifx\csname urlstyle\endcsname\relax
  \providecommand{\doi}[1]{DOI~\discretionary{}{}{}#1}\else
  \providecommand{\doi}{DOI~\discretionary{}{}{}\begingroup
  \urlstyle{rm}\Url}\fi
\providecommand{\eprint}[2][]{\url{#2}}

\bibitem[{Abbod et~al.(2007)Abbod, Sellars, Cizek, Linkens, and
  Mahfouf}]{abbod2007modeling}
Abbod MF, Sellars CM, Cizek P, Linkens DA, Mahfouf M (2007) {Modeling the flow
  behavior, recrystallization, and crystallographic texture in hot-deformed
  Fe-30 Wt Pct Ni Austenite}. Metall Mater Trans A 38(10):2400--2409

\bibitem[{Acharya(2010)}]{acharya2010new}
Acharya A (2010) New inroads in an old subject: plasticity, from around the
  atomic to the macroscopic scale. J Mech Phys Solids 58(5):766--778

\bibitem[{Anand et~al.(2015)Anand, Gurtin, and Reddy}]{anand2015stored}
Anand L, Gurtin ME, Reddy BD (2015) The stored energy of cold work, thermal
  annealing, and other thermodynamic issues in single crystal plasticity at
  small length scales. Int J Plasticity 64:1--25

\bibitem[{Ayers(1994)}]{ayers1994measurment}  
Ayers JE (1994) The measurement of threading dislocation densities in semiconductor crystals by X-ray diffraction. J Crystal Growth 135(1-2):71--77.

\bibitem[{Berdichevsky(2005)}]{berdichevsky2005homogenization}
Berdichevsky VL (2005) Homogenization in micro-plasticity. J
  Mech Phys Solids 53(11):2457--2469

\bibitem[{Berdichevsky(2006{\natexlab{a}})}]{berdichevsky2006continuum}
Berdichevsky VL (2006{\natexlab{a}}) Continuum theory of dislocations
  revisited. Contin Mech Thermodyn 18(3-4):195--222

\bibitem[{Berdichevsky(2006{\natexlab{b}})}]{berdichevsky2006thermodynamics}
Berdichevsky VL (2006{\natexlab{b}}) On thermodynamics of crystal plasticity.
  Scripta Mater 54(5):711--716

\bibitem[{Berdichevsky(2008)}]{berdichevsky2008entropy}
Berdichevsky VL (2008) Entropy of microstructure. J Mech Phys Solids 56(3):742--771

\bibitem[{Berdichevsky(2019)}]{berdichevsky2019beyond}
Berdichevsky VL (2019) Beyond classical thermodynamics: Dislocation-mediated
  plasticity. J Mech Phys Solids 129:83--118

\bibitem[{Berdichevsky and Le(2007)}]{berdichevsky2007dislocation}
Berdichevsky VL, Le KC (2007) Dislocation nucleation and work hardening in
  anti-plane constrained shear. Contin Mech Thermodyn
  18(7-8):455--467

\bibitem[{Berdichevsky and Sedov(1967)}]{berdichevsky1967dynamic}
Berdichevsky VL, Sedov LI (1967) Dynamic theory of continuously distributed
  dislocations. its relation to plasticity theory. J Appl
  Math Mech 31(6):989--1006

\bibitem[{Bilby(1955)}]{bilby1955types}
Bilby B (1955) Types of dislocation source. In: Report of Bristol Conference on
  Defects in Crystalline Solids (Bristol 1954, London: The Physical Soc.), pp
  124--133

\bibitem[{Calcagnotto et~al.(2010)Calcagnotto, Ponge, Demir, and
  Raabe}]{calcagnotto2010orientation}
Calcagnotto M, Ponge D, Demir E, Raabe D (2010) Orientation gradients and
  geometrically necessary dislocations in ultrafine grained dual-phase steels
  studied by 2d and 3d ebsd. Mater Sci Eng A
  527(10-11):2738--2746

\bibitem[{Follansbee and Kocks(1988)}]{follansbee1988constitutive}
Follansbee PS, Kocks UF (1988) A constitutive description of the deformation of
  copper based on the use of the mechanical threshold stress as an internal
  state variable. Acta Metall 36(1):81--93

\bibitem[{Groma et~al.(2003)Groma, Csikor, and Zaiser}]{groma2003spatial}
Groma I, Csikor F, Zaiser M (2003) Spatial correlations and higher-order
  gradient terms in a continuum description of dislocation dynamics. Acta
  Mater 51(5):1271--1281

\bibitem[{Hochrainer(2016)}]{hochrainer2016thermodynamically}
Hochrainer T (2016) Thermodynamically consistent continuum dislocation
  dynamics. J Mech Phys Solids 88:12--22

\bibitem[{Kr{\"o}ner(1955)}]{kroner1955fundamentale}
Kr{\"o}ner E (1955) Der fundamentale zusammenhang zwischen versetzungsdichte
  und spannungsfunktionen. Z Phys 142(4):463--475

\bibitem[{Kr{\"o}ner(1958)}]{kroner1958kontinuumstheorie}
Kr{\"o}ner E (1958) Kontinuumstheorie der Versetzungen und Eigenspannungen,
  vol~5. Springer

\bibitem[{Langer(2015)}]{langer2015}
Langer JS (2016) Statistical thermodynamics of strain hardening in polycrystalline solids. Phys Rev E 92(3):032125

\bibitem[{Langer and Le(2020)}]{langer2020scaling}
Langer JS, Le KC (2020) Scaling confirmation of the thermodynamic dislocation
  theory. Proc Natl Acad Sci USA 117(47):29431-29434

\bibitem[{Langer et~al.(2010)Langer, Bouchbinder, and
  Lookman}]{langer2010thermodynamic}
Langer JS, Bouchbinder E, Lookman T (2010) Thermodynamic theory of
  dislocation-mediated plasticity. Acta Mater 58(10):3718--3732

\bibitem[{Le(2018)}]{le2018athermodynamic}
Le KC (2018) Thermodynamic dislocation theory for non-uniform plastic
  deformations. J Mech Phys Solids 111:157--169

\bibitem[{Le(2020a)}]{le2020two}
Le KC (2020a) Two universal laws for plastic flows and the consistent
  thermodynamic dislocation theory. Mech Res Commun 109:103597
  
\bibitem[{Le(2020b)}]{le2020introduction}
Le KC (2020b) Introduction to Micromechanics. Nova Science, New York.

\bibitem[{Le and Piao(2019)}]{le2019thermodynamic}
Le KC, Piao Y (2019) Thermodynamic dislocation theory: Size effect in torsion.
  Int J Plasticity 115:56--70

\bibitem[{Le and Sembiring(2008)}]{le2008analytical}
Le KC, Sembiring P (2008) Analytical solution of plane constrained shear
  problem for single crystals within continuum dislocation theory. Arch Appl
  Mech 78(8):587--597

\bibitem[{Le and Stumpf(1996)}]{le1996model}
Le KC, Stumpf H (1996) A model of elastoplastic bodies with continuously
  distributed dislocations. Int J Plasticity 12(5):611--627

\bibitem[{Le and Tran(2018)}]{le2018cthermodynamic}
Le KC, Tran TM (2018) Thermodynamic dislocation theory: Bauschinger effect.
  Phys Rev E 97(4):043002

\bibitem[{Le et~al.(2017)Le, Tran, and Langer}]{le2017thermodynamic}
Le KC, Tran TM, Langer JS (2017) Thermodynamic dislocation theory of
  high-temperature deformation in aluminum and steel. Phys Rev E 96:013004

\bibitem[{Le et~al.(2018)Le, Tran, and Langer}]{le2018thermodynamic}
Le KC, Tran TM, Langer JS (2018) Thermodynamic dislocation theory of adiabatic
  shear banding in steel. Scripta Mater 149:62--65

\bibitem[{Levitas and Javanbakht(2015)}]{levitas2015thermodynamically}
Levitas VI, Javanbakht M (2015) Thermodynamically consistent phase field
  approach to dislocation evolution at small and large strains. J Mech Phys Solids 82:345--366

\bibitem[{Lieou and Bronkhorst(2020)}]{lieou2020thermodynamic}
Lieou CK, Bronkhorst CA (2020) Thermodynamic theory of crystal plasticity:
  formulation and application to polycrystal fcc copper. J Mech Phys Solids  103905

\bibitem[{Marchand and Duffy(1988)}]{marchand1988experimental}
Marchand A, Duffy J (1988) An experimental study of the formation process of
  adiabatic shear bands in a structural steel. J Mech Phys Solids
  36(3):251--283

\bibitem[{Morito et~al.(2003)}]{morito2003dislocation}  
Morito, S., Nishikawa, J. and Maki, T. (2003) Dislocation density within lath martensite in Fe-C and Fe-Ni alloys. ISIJ Int 43(9):1475--1477.

\bibitem[{Mura(1965)}]{mura1965continuous}
Mura T (1965) Continuous distribution of dislocations and the mathematical
  theory of plasticity. Phys Status Solidi B 10(2):447--453

\bibitem[{Nye(1953)}]{nye1953some}
Nye J (1953) Some geometrical relations in dislocated crystals. Acta
  Metall 1(2):153--162

\bibitem[{Ortiz and Repetto(1999)}]{ortiz1999nonconvex}
Ortiz M, Repetto E (1999) Nonconvex energy minimization and dislocation
  structures in ductile single crystals. J Mech Phys Solids 47(2):397--462

\bibitem[{Po et~al.(2019)Po, Huang, and Ghoniem}]{po2019continuum}
Po G, Huang Y, Ghoniem N (2019) A continuum dislocation-based model of wedge
  microindentation of single crystals. Int J Plasticity 114:72--86

\bibitem[{Samanta(1971)}]{samanta1971dynamic}
Samanta SK (1971) Dynamic deformation of aluminium and copper at elevated
  temperatures. J Mech Phys Solids 19(3):117--135

\bibitem[{Shi et~al.(1997)Shi, McLaren, Sellars, Shahani, and
  Bolingbroke}]{shi1997constitutive}
Shi H, McLaren AJ, Sellars CM, Shahani R, Bolingbroke R (1997) Constitutive
  equations for high temperature flow stress of aluminium alloys. Mater Sci
  Technol 13(3):210--216

\bibitem[{Weertman(1996)}]{weertman1996dislocation}
Weertman JH (1996) Dislocation based fracture mechanics. World Scientific
  Publishing Company

\end{thebibliography}
\end{document}